\documentclass[sigplan,screen]{acmart}

\usepackage{natbib}
\usepackage{pifont}
\usepackage{soul}
\usepackage{tikz}
\usepackage{amsmath}
\usepackage{comment}
\usepackage{tabularx}
\usepackage{color,soul}

\usepackage{graphics}
\usepackage{cleveref}
\usepackage{tikz}
\usepackage{subcaption}
\usepackage{pgfplots}
\pgfplotsset{compat=1.16}
\pgfplotsset{hide scale/.style={
/pgfplots/xtick scale label code/.code={},
/pgfplots/ytick scale label code/.code={}}}

\usepackage{listings}

\definecolor{codegreen}{rgb}{0,0.6,0}
\definecolor{codegray}{rgb}{0.5,0.5,0.5}
\definecolor{codepurple}{rgb}{0.58,0,0.82}
\definecolor{backcolour}{rgb}{0.95,0.95,0.92}

\lstdefinelanguage{mystyle}{
    morekeywords=[2]{void, for, if, int, assert, break, continue, return, const, goto, do, while, sizeof, bool},
    morekeywords=[3]{nullptr, ESTALE, true, false, NUM\_SLOTS, static},
    morekeywords=[4]{GlobalAgent, Agent, Thread, Cpu, Transaction, map, vector, std, auto, uint64_t, mmio\_queue, message},
}

\lstdefinestyle{mystyle}{
    language={mystyle},
    commentstyle=\color{codegreen},
    numberstyle=\tiny\color{codegray},
    stringstyle=\color{codepurple},
    basicstyle=\ttfamily\footnotesize,
    breakatwhitespace=false,         
    breaklines=true,                 
    captionpos=b,                    
    keepspaces=true,                 
    numbers=left,                    
    numbersep=5pt,                  
    showspaces=false,                
    showstringspaces=false,
    showtabs=false,                  
    tabsize=2,
    morecomment=[l]{//},
    keywordstyle=[2]{\color{magenta}},
    keywordstyle=[3]{\color{red}},
    keywordstyle=[4]{\color{blue}},
}

\usepackage{xspace}
\newcommand{\name}{{Wave}\xspace} % If you change the name to something that starts with a vowel, be sure you replace "a" with "an" throughout the paper.
\usepackage[normalem]{ulem}

\usepackage{lipsum}

\copyrightyear{2025}
\acmYear{2025}
\setcopyright{cc}
\setcctype{by}
\acmConference[ASPLOS '25]{Proceedings of the 30th ACM International Conference on Architectural Support for Programming Languages and Operating Systems, Volume 3}{March 30-April 3, 2025}{Rotterdam, Netherlands}
\acmBooktitle{Proceedings of the 30th ACM International Conference on Architectural Support for Programming Languages and Operating Systems, Volume 3 (ASPLOS '25), March 30-April 3, 2025, Rotterdam, Netherlands}\acmDOI{10.1145/3676642.3736113}
\acmISBN{979-8-4007-1080-3/2025/03}

\keywords{Operating systems, SmartNICs}

\ccsdesc[500]{Computer systems organization}

\begin{document}

\title{\name: Offloading Resource Management\\to SmartNIC Cores}

\author{Jack Tigar Humphries}
\authornote{Work completed while at Google, Inc. and Stanford University.}
\email{jhumphri@cs.stanford.edu}
\affiliation{%
  \institution{Stellar Development Foundation}
  \city{San Francisco, CA}
  \country{USA}}

\author{Neel Natu}
\email{neelnatu@google.com}
\affiliation{%
  \institution{Google, Inc.}
  \city{Sunnyvale, CA}
  \country{USA}}

\author{Kostis Kaffes}
\email{kkaffes@cs.columbia.edu}
\affiliation{%
  \institution{Columbia University}
  \city{New York, NY}
  \country{USA}}

\author{Stanko Novakovi\'c}
\email{stanko@google.com}
\affiliation{%
  \institution{Google, Inc.}
  \city{Seattle, WA}
  \country{USA}}

\author{Paul Turner}
\email{pjt@google.com}
\affiliation{%
  \institution{Google, Inc.}
  \city{Sunnyvale, CA}
  \country{USA}}

\author{Henry M. Levy}
\email{hanklevy@google.com}
\affiliation{%
  \institution{Google, Inc.}
  \city{Seattle, WA}
  \country{USA}}
  \affiliation{%
  \institution{University of Washington}
  \city{Seattle, WA}
  \country{USA}}

\author{David Culler}
\email{dculler@google.com}
\affiliation{%
  \institution{Google, Inc.}
  \city{Sunnyvale, CA}
  \country{USA}}
  \affiliation{%
  \institution{University of California, Berkeley}
  \city{Berkeley, CA}
  \country{USA}}

\author{Christos Kozyrakis}
\email{kozyraki@stanford.edu}
\affiliation{%
  \institution{Stanford University}
  \city{Stanford, CA}
  \country{USA}}

% We don't want the superscripts for affiliations to appear in the ACM reference format. Plus, there is an "and" included before the last author in the reference format but the template thinks we only have one author due to the way we did it above. Thus, redo the authors the way they should appear in the ACM reference format.
\renewcommand{\authors}{Jack Tigar Humphries, Neel Natu, Kostis Kaffes, Stanko Novakovi\'c, Paul Turner, Henry M. Levy, David Culler, and Christos Kozyrakis}
\renewcommand{\shortauthors}{Jack Tigar Humphries et al.}

\date{}

\begin{abstract}
\label{sec:abstract}

SmartNICs are increasingly deployed in datacenters to offload tasks from server CPUs, improving the efficiency and flexibility of datacenter security, networking and storage. 
Optimizing cloud server efficiency in this way is critically important to ensure that virtually all server resources are available to paying customers.
Userspace system software, specifically, decision-making tasks performed by various operating system subsystems, is particularly well suited for execution on mid-tier SmartNIC ARM cores.
To this end, we introduce \name, a framework for offloading userspace system software to processes/agents running on the SmartNIC.
\name uses Linux userspace systems to better align system functionality with SmartNIC capabilities.
It also introduces a new host-SmartNIC communication API that enables offloading of even $\mu$s-scale system software. 
To evaluate \name, we offloaded preexisting userspace system software including kernel thread scheduling, memory management, and an RPC stack to SmartNIC ARM cores, which showed a performance degradation of 1.1\%-7.4\% in an apples-to-apples comparison with on-host implementations.
\name recovered host resources consumed by on-host system software for memory management (saving 16 host cores), RPCs (saving 8 host cores), and virtual machines (an 11.2\% performance improvement).
\name highlights the potential for rethinking system software placement in modern datacenters, unlocking new opportunities for efficiency and scalability.

\end{abstract}

\maketitle

\section{Introduction}
\label{sec:introduction}
Cloud providers continuously optimize server efficiency to ensure that virtually all server resources are available to paying customers.
They do this by offloading virtualization and network stacks to accelerators on SmartNICs and other specialized offload devices, such as the AWS Nitro~\cite{nitro}, NVIDIA BlueField~\cite{bluefield}, Fungible DPU~\cite{fungible}, AMD Pensando DPU~\cite{pensando}, and Intel Mount Evans~\cite{intel-mount-evans}.
For example, AWS offloads virtualized I/O and cloud security to their Nitro device to improve host utilization and reduce jitter.
Recent work~\cite{dagger, flextoe, 1rma, erss, azure-accelerated-networking, lynx, nanopu} offloads the host operating system network data plane, freeing host compute resources and enabling acceleration.

Although some of these benefits could be realized with additional on-host compute, cloud providers deploy SmartNICs in many new datacenter machines because they offer the simplest and cheapest way to provide a uniform security and management view across machines regardless of vendor, architecture type, and maturity/availability of software~\cite{aws-nitro-hotchips-2019}.
Cloud providers also have tighter control over SmartNIC design than host architecture.
They purchase large volumes of network hardware, so they work closely with hardware vendors to
expedite decision-making~\cite{intel-mount-evans-joint} about accelerator logic, memory size, core count, and more.
In contrast, host CPU architectures are often designed by external vendors for a wide range of cases and customers, so cloud providers cannot tailor these architectures to their specific needs.

Beyond acceleration, cloud providers also deploy mid-tier ARM cores inside SmartNICs, such as 16 ARM cores in Intel Mount Evans~\cite{intel-mount-evans}.
Although networking and storage tasks for virtual machines are being successfully offloaded to SmartNIC accelerators and datapaths, the systems community is still exploring ways to fully utilize these mid-tier  ARM cores~\cite{characterizing-off-path-smartnic, the-rise-of-smartnics, on-disadvantages-of-programmable-nics, azure-accelerated-networking, mind-the-gap}.
Offloaded applications suffer from the low compute performance of ARM cores and high host-SmartNIC communication latency, occasionally experiencing worse end-to-end performance even with the additional SmartNIC resources~\cite{mind-the-gap}.
Additionally, the complexity~\cite{clara, lognic, floem} involved in offloading applications across PCIe likely makes it impractical on a datacenter-wide basis, and, crucially, security concerns prevent many potential uses: providers typically restrict these ARM cores from running arbitrary code~\cite{intel-mount-evans-venture-beat, no-aws-operator-access}.

Userspace system software is particularly well-suited for execution on SmartNIC ARM cores.
This software consists of the \textit{decision-making tasks performed by various subsystems}—such as choosing the next thread to schedule or determining the physical memory frame to allocate for a virtual page.
Because system software is deployed uniformly across all datacenter machines, the effort required to offload it is invested only once, subsequently providing consistent benefits across the entire infrastructure.
Moreover, significant existing work on microkernels~\cite{l4-microkernel-lessons, redleaf, l3-to-sel4, microkernel-goes-general, okl4, qnx, theseus, xpc, fuschia-zircon, harmonizing-microkernels, minix3, pikeos, sel4, hydra, improving-ipc-by-kernel-design, on-microkernel-construction, filesystem-semi-microkernel, mach}, multikernels~\cite{barrelfish, barrelfish-dc, barrelfish-intel-single-chip, barrelfish-cosh, barrelfish-msr, menzi-masters-thesis, barrelfish-technical-note-001, barrelfish-technical-note-005, nros}, exokernels~\cite{exokernel}, and Linux userspace subsystems~\cite{ix, shinjuku, shenango, reflex, zygos, caladan, snap, persephone, demikernel, dune, ghost, syrup, ccuserspace, userfaultfd, fuse, terra-incognita, uio, apple-driverkit, windows-umdf, pond.asplos23} has already established the necessary separation of such systems, simplifying this transition.
ARM cores align naturally with this decision-making since decisions typically involve straightforward heuristics, offering a good balance between power efficiency and computational demand.
Finally, because cloud providers maintain full control over the host system software, this software is inherently trusted to operate within privileged execution environments on SmartNICs.

The main challenge in offloading system software is overcoming the high PCIe interconnect latency in a manner that works for a wide range of software with different requirements.
Thread scheduling and RPC stacks require sub-$\mu$s communication~\cite{nsight, ghost} as cores remain idle until policy decisions are enforced; 
memory management requires high-throughput communication to move an entire address space of page table entries from the host to the SmartNIC and a decision for each migrated page back to the host.
PCIe presents a distinct challenge because microkernels, some multikernels, and Linux userspace subsystems like ghOSt~\cite{ghost} and Snap~\cite{snap} target coherent CPUs. Even those adapted for non-coherent CPUs~\cite{barrelfish-intel-single-chip, barrelfish-cosh, barrelfish-msr, menzi-masters-thesis, barrelfish-technical-note-001, barrelfish-technical-note-005} benefit from the much lower latencies of the on-chip  CPU interconnect. 
The higher PCIe latency arises from packet-based protocols, flow control, bridge logic, and lower clock rates compared to internal CPU interconnects.
Previous approaches, such as Floem~\cite{floem} and iPipe~\cite{ipipe}, optimize primarily for throughput using DMA-based queues, which do not satisfy the strict latency and reliability guarantees essential for system software.

We introduce \name, a framework for offloading system software to the SmartNIC, and use \name to offload three preexisting system software components.
\name provides a general mechanism for host-SmartNIC communication that enables offloading of even $\mu$s-scale system software.

Our first key insight is that \textit{offloading system software to a SmartNIC requires the same API and guarantees that ghOSt~\cite{ghost} provides for running system software in userspace.}
The ghOSt API supports atomic commits of policy decisions made by userspace software.
This strong guarantee becomes even more essential when the userspace system software that makes decisions operates across a high-latency PCIe interconnect.
\name implements this API with a unidirectional queue for sending messages from the host to the SmartNIC and another unidirectional queue for sending decisions back.

Our next key insight is that \textit{different system software patterns must use different memory mechanisms that provide different performance guarantees. }
\name leverages DMA for high-throughput transfers, which is required by memory management, and MMIO for low-latency communication, which is needed by the thread scheduler and the RPC stack.
Even with MMIO, \name uses different types of page table entries (write-combining and write-through) for different data types.
It uses write-combining entries to further reduce the latency of the host-to-NIC queue with a batching optimization and write-through entries to hide the latency of the NIC-to-host queue with prefetching and caching optimizations.
\name also precomputes and prestages decisions to hide the remaining host-SmartNIC communication latency.

As in ghOSt, \name implements system software in userspace \textit{agents} running on the SmartNIC across the PCIe interconnect.
We prototype \name on Intel Mount Evans~\cite{intel-mount-evans} and demonstrate practical offload of preexisting system software, including kernel thread scheduling, memory management, and an RPC stack.

We show that  \name communication/notification mechanisms are sufficiently fast (426ns for an agent on the SmartNIC to write a thread scheduling decision and send an interrupt) to make it practical to offload system software to SmartNIC ARM cores.
We further show that \name's optimizations are essential to hiding the PCIe latency, improving the RocksDB~\cite{rocksdb} key-value store throughput by $\sim$350\% compared to an offloaded system with no optimizations.
These optimizations make offload practical, with a small performance degradation of 1.1-7.4\% compared to an on-host implementation.
Finally, we show that \name enables practical offload of compute-heavy software, such as machine-learning-based memory management software, which reduces RocksDB's memory footprint by 79\%.
This compute-heavy software can now leverage SmartNIC compute, saving 16 host cores.
We also use \name to offload an RPC stack, freeing 8 host cores, and reduce interference with virtual machines, improving their turbo performance by 11.2\%.

In summary, our key contributions include:
\begin{itemize}
\item Transparently offloading three preexisting userspace system software components with full compatibility with the Linux kernel and unmodified applications.
\item Building a host-SmartNIC communication API that uses diverse mechanisms to manage the PCIe latency gap and enable $\mu$s-scale offloaded system software.
\item Providing apples-to-apples comparisons of offloaded vs. on-host system software components that show small offload performance degradation of 1.1-7.4\%.
\item Saving host resources with offload, e.g., recovering 16 host cores from memory management and 8 from RPC.
\end{itemize}
\section{Offloading System Software}
\label{sec:why-offload-the-os}
\subsection{SmartNICs and IPU Background}
\label{sec:background:smartnics-and-SmartNICs}

\textbf{SmartNIC Capabilities.} SmartNICs are PCIe cards featuring a System-on-Chip (SoC), which consists of a network interface card, compute capabilities in the form of ARM CPUs or programmable P4 pipelines, and hardware accelerators for compression, encryption, RDMA, NVMe virtualization, NVMe-over-fabrics, hardware clock synchronization, etc.~\cite{nitro,bluefield, intel-mount-evans}.
The interface between the host and currently available SmartNICs is limited to memory-mapped IO (MMIO), direct memory access (DMA), and interrupts.
The host can access the SoC's DRAM via MMIO operations, which are synchronous memory operations over the PCIe interface that take on the order of 1$\mu$s.
DMA allows the SmartNIC to place data in the host memory without involving the host CPU, and vice versa.
Finally, the SmartNIC can send MSI-X signals (interrupts) to host cores.
We analyze the overheads of these operations in our setup in \S\ref{sec:evaluation:microbenchmarks}.

\textbf{Offloads.} Recent work offloads infrastructure functionality to SmartNIC accelerators to leverage their computational power, turning SmartNICs into the \textit{Infrastructure Processing Units (IPUs)} of the datacenter.
The AWS Nitro~\cite{nitro} accelerates virtual machine control planes and data planes.
AccelNet~\cite{azure-accelerated-networking}, Dagger~\cite{dagger}, FlexTOE~\cite{flextoe}, 1RMA~\cite{1rma}, and eRSS~\cite{erss} offload a host OS network stack data plane to the SmartNIC.

However, significant barriers prevent utilization of  SmartNIC ARM CPUs.
Prior work \cite{characterizing-off-path-smartnic, the-rise-of-smartnics, on-disadvantages-of-programmable-nics, azure-accelerated-networking} identifies the weak performance of the general-purpose ARM cores relative to the host as a significant obstacle to offload.
Floem~\cite{floem} and Mind the Gap~\cite{mind-the-gap} demonstrate that host-SmartNIC communication via the slow PCIe interconnect further degrades end-to-end workload performance--in some cases performing worse than with no offload at all--and propose optimizations.
Beyond this, offloading workloads piecemeal is not a cohesive vision for the entire datacenter since individual workloads often run on only a subset of the machines.

\subsection{Why System Software?}
System software is a natural choice for offloading to the SmartNIC ARM cores.
It requires general-purpose compute to support a wide range of workloads and hardware, with flexibility being paramount.
It parallelizes easily with abstractions that can be shared or pipelined across several cores, such as address spaces in a memory manager and NUMA domains (e.g., chiplets and sockets) in a thread scheduler.
System software reduces its interference with on-host workloads by running on the SmartNIC and leverages the unique network insight that a SmartNIC provides to optimize the entire machine.
Because every server in the fleet runs system software, each can offload its system software to a SmartNIC, ensuring that any offload benefits apply \textit{universally}.

\subsection{How Is Split OS Decision-Making Accomplished?}

Offloading system software to the SmartNIC requires the decision-making logic to be separated from the host kernel.
Fortunately, a wealth of prior work has already established or leveraged such separation, moving OS decision-making to separate userspace processes, so much of the required structure is already in place.
Microkernels~\cite{l4-microkernel-lessons, redleaf, l3-to-sel4, microkernel-goes-general, okl4, qnx, theseus, xpc, fuschia-zircon, harmonizing-microkernels, minix3, pikeos, sel4, hydra, improving-ipc-by-kernel-design, on-microkernel-construction, filesystem-semi-microkernel, mach}, multikernels~\cite{barrelfish, barrelfish-dc, barrelfish-intel-single-chip, barrelfish-cosh, barrelfish-msr, menzi-masters-thesis, barrelfish-technical-note-001, barrelfish-technical-note-005, nros}, exokernels~\cite{exokernel}, and userspace resource management systems~\cite{ix, shinjuku, shenango, reflex, zygos, caladan, snap, persephone, demikernel, dune, ghost, syrup, ccuserspace, userfaultfd, fuse, terra-incognita, uio, apple-driverkit, windows-umdf, pond.asplos23} all illustrate that OS functionality can be effectively split out to run in userspace provided there are robust and efficient mechanisms to communicate with the kernel.
Even Linux, a monolithic kernel, includes partial decoupling of certain services—enabling, for example, user-level drivers and specialized control planes—so additional invasive changes are not necessary.

\textbf{Microkernels.} Microkernels~\cite{l4-microkernel-lessons, redleaf, l3-to-sel4, microkernel-goes-general, okl4, qnx, theseus, xpc, fuschia-zircon, harmonizing-microkernels, minix3, pikeos, sel4, hydra, improving-ipc-by-kernel-design, on-microkernel-construction, filesystem-semi-microkernel, mach} offload OS services from the kernel to userspace processes, keeping only minimal functionality inside the kernel.
Significantly, they emphasize optimized inter-process communication mechanisms since slow IPC in the OS-userspace critical path has traditionally been the drawback of these systems.
Several major proposals have led to order of magnitude performance improvements~\cite{l4-microkernel-lessons, l3-to-sel4}, including asynchronous communication; message-passing with a single copy operation; sending a \textit{batch} of several messages; and passing messages to "port-like" IPC endpoint handlers rather than directly to threads.

\textbf{Multikernels.} Multikernels~\cite{barrelfish, barrelfish-dc, barrelfish-intel-single-chip, barrelfish-cosh, barrelfish-msr, menzi-masters-thesis, barrelfish-technical-note-001, barrelfish-technical-note-005, nros} like Barrelfish~\cite{barrelfish, barrelfish-dc, barrelfish-intel-single-chip, barrelfish-cosh, barrelfish-msr, menzi-masters-thesis, barrelfish-technical-note-001, barrelfish-technical-note-005} propose a new OS design to work across CPUs with many cores, varying cache coherence protocols (or none at all), memory hierarchies, architectures, performance characteristics, and I/O configurations.
Multikernels particularly aim to perform well on machines with large numbers of cores and so place one OS image on each core, with each core operating independently.
Cores do not share memory with each other to limit coherence traffic and improve scalability.
They communicate by passing messages via highly optimized channels, using techniques such as batching, pipelining, asynchrony, and cache-line alignment, taking care to avoid TLB flushes and cache pollution.
Barrelfish is implemented on both cache-coherent~\cite{barrelfish, barrelfish-dc} and non-coherent machines~\cite{barrelfish-intel-single-chip, barrelfish-cosh, barrelfish-msr, menzi-masters-thesis, barrelfish-technical-note-001, barrelfish-technical-note-005}.

\textbf{Exokernel.} Exokernel ~\cite{exokernel} exports hardware resources to untrusted userspace applications.
Applications implement their own OS services to improve performance.
While Exokernel proposes optimizations like asynchronous IPC that Wave can exploit, its focus is largely orthogonal to ours.

\textbf{Userspace Resource Management.} Recent userspace resource management systems bypass the kernel and its associated overheads and scaling constraints, instead implementing their own custom control and data planes in userspace~\cite{ix, shinjuku, shenango, reflex, zygos, caladan, snap, persephone, demikernel, dune, ghost, syrup, ccuserspace, userfaultfd, fuse, terra-incognita, uio, apple-driverkit, windows-umdf, pond.asplos23}.
Workloads running on these systems perform significantly better than on vanilla Linux, e.g., 3.6x better throughput and a 2x latency reduction in IX~\cite{ix}.
However, they sacrifice compatibility and require developers to re-architect and port their applications.

As a middle ground, a recent trend offloads to userspace only the control plane of specific subsystems and keeps the data plane in the kernel.
This approach maintains compatibility and does not sacrifice performance, while letting users easily specialize and fine-tune policies to their needs.
ghOSt~\cite{ghost} and Syrup~\cite{syrup} offload the kernel thread scheduling control plane to userspace. They make it easy to implement custom policies and offer benefits such as rebootless upgrades, fault isolation, and state encapsulation.
CCP~\cite{ccuserspace} proposes offloading the congestion control plane to a userspace agent to accelerate experimentation.
userfaultfd~\cite{userfaultfd} offloads the page fault handler to userspace applications, which is used to support live migration of virtual machines~\cite{userfaultfd-vm}, while PageFlex~\cite{yelam25pageflex} offloads only the reclamation policy. Userspace tools also have access to interfaces for scanning access and dirty page table bits, which can be used by memory policy decisions~\cite{pond.asplos23}.
FUSE~\cite{fuse}, which lets a userspace application expose a file system to the Linux kernel, has been used to implement over 100 filesystems ~\cite{terra-incognita}.
UIO~\cite{uio}, Apple DriverKit~\cite{apple-driverkit}, UMDF~\cite{windows-umdf} and the VFIO driver framework~\cite{vfio} support userspace drivers.

Each system has its own ad hoc scheme for kernel-userspace state sharing and synchronization to prevent time-of-check to time-of-use bugs.
State includes network streams, thread runnability, and page table entries.
ghOSt uses queues and message barriers and userfaultfd uses file descriptors with blocking syscalls.
All communicate over coherent shared memory, so they can use a single queue API, though their performance would degrade over PCIe.

\subsection{Why offload to a SmartNIC rather than deploy a heterogeneous host CPU architecture?}
\label{sec:background:why-an-SmartNIC}

Moving system software onto a SmartNIC might appear counterintuitive given the existence of heterogeneous architectures with efficiency cores, such as ARM big.LITTLE~\cite{arm-big-little} and Apple M-series efficiency cores~\cite{apple-m-series}.
Some software, such as thread scheduling, consumes $\sim$5\% of host cycles~\cite{profiling-a-warehouse-scale-computer}, so why not dedicate 5 host efficiency cores out of 100+ to the software?
One could imagine assigning userspace system software like the Shinjuku scheduler~\cite{shinjuku} or the ghOSt global agent~\cite{ghost} to dedicated efficiency cores, leaving high-performance cores exclusively for application workloads.

However, this approach is impractical because heterogeneous CPU architectures are not widely available in datacenters. 
Conversely, SmartNICs are deployed in many new datacenter machines since they offer a simple, cost-effective way to provide uniform security and management across diverse hardware from different vendors, architectures, and software maturities.
The SmartNIC thus provides a consistent, universal platform for offloading system software, making it the logical and strategic choice.
Additionally, since cloud providers control the design of SmartNICs~\cite{google-ipu}, specialized accelerators tailored to their workloads can be integrated directly into the SmartNIC SoC.
This enables tighter hardware-software co-design between I/O operations and system software.
Furthermore, since all I/O events pass through SmartNICs in cloud deployments, leveraging this architecture for system software is both natural and advantageous.
\section{\name Design}
\label{sec:wave-design}
\name is a framework for offloading userspace system software to processes/agents running on the SmartNIC.
The in-kernel mechanisms that such software relies on remain on the host, which invokes the SmartNIC software over PCIe.

\subsection{\name Overview}
\label{sec:wave-design:overview}

\begin{figure}[t]
\centering
\includegraphics[width=0.75\columnwidth,keepaspectratio]{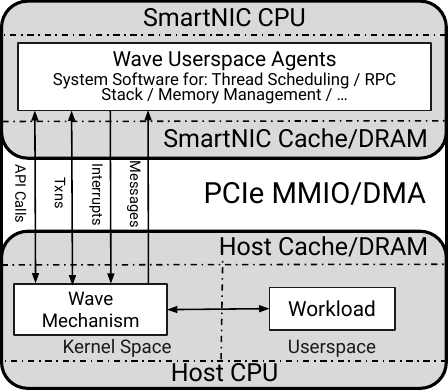}
\caption{High-level \name design.}
\Description[High-level Wave design.]{This shows the high-level design of Wave.}
\label{fig:wave-in-a-nutshell}
\end{figure}

\Cref{fig:wave-in-a-nutshell} shows the high-level \name design.
\name implements a shared memory queue abstraction backed by a choice of MMIO or DMA over the PCIe interconnect.
The host kernel uses the queue to send the required state to the \name agents, which run the software to make resource allocation decisions.
Each agent can run a single system software component or  combine software if beneficial.
One such example is combining the RPC stack and thread scheduling in \S\ref{sec:evaluation:networking}.
The agents send their decisions back to the host kernel via a transaction-based API backed by another queue.
The host kernel then enforces the decisions. \name enables existing agent implementations, such as ghOSt agents~\cite{ghost}, memory managers~\cite{userfaultfd}, and network agents~\cite{snap}, to run on the SmartNIC and enforce decisions on the host.

% https://docs.google.com/drawings/d/1a0yB8BgveHCjkHfTjglLBmtHkkSuQpyjyAF6JXaFJZ8/edit

\begin{figure*}[t]
\centering
\includegraphics[width=0.9\textwidth,height=\textheight,keepaspectratio]{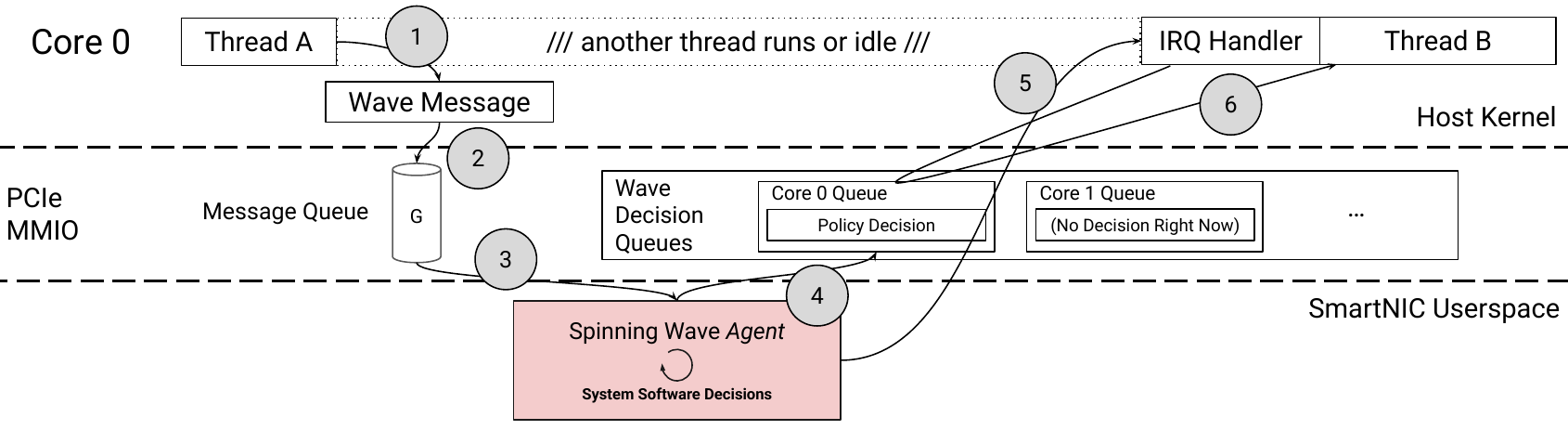}
\caption{\name schedules host cores across the PCIe interconnect.}
\Description[A scheduler in Wave.]{Wave schedules host cores across the PCIe interconnect.}
\label{fig:wave-detailed}
\end{figure*}
% https://docs.google.com/drawings/d/1cXmUVJIYjfdmqzGY_Bi7AjC2rgZwp0mNYqliwKv8MkQ/edit

We now explain how \name makes decisions using thread scheduling as an example. \Cref{fig:wave-detailed} shows the lifetime of a \name scheduling decision.
A global polling agent model is used, though \name scales to multiple agents~\cite{ghost}.

\ding{182} When Thread A triggers an event (e.g., blocking on a futex, allocating a virtual memory region, etc.) on host CPU core 0, the host sends a message to the agent on the SmartNIC.
\ding{183} The message is added to a message queue, which is backed by the SmartNIC DRAM and accessible via the \name shared memory abstraction.
\ding{184} The agent, which runs in userspace on the SmartNIC CPU, polls for messages from the queue.
\name system software polls because decisions such as thread scheduling require low latency.
Upon reading the message, the agent runs the system software to determine how to allocate resources for that subsystem.
\ding{185} The agent writes its decision to the decision queue in SmartNIC memory.
\ding{186} The agent sends an interrupt (MSI-X) to core 0 on the host to trigger a preemption.
\ding{187} In the interrupt handler, the host kernel on core 0 reads the decision via the \name shared memory queue, commits it atomically with a transaction, and enforces it, e.g., by context switching to thread B.

Compared with existing solutions that offload system software to userspace~\cite{l4-microkernel-lessons, redleaf, l3-to-sel4, microkernel-goes-general, okl4, qnx, theseus, xpc, fuschia-zircon, harmonizing-microkernels, minix3, pikeos, sel4, hydra, improving-ipc-by-kernel-design, on-microkernel-construction, filesystem-semi-microkernel, mach, barrelfish, barrelfish-dc, barrelfish-intel-single-chip, barrelfish-cosh, barrelfish-msr, menzi-masters-thesis, barrelfish-technical-note-001, barrelfish-technical-note-005, exokernel, ix, shinjuku, shenango, reflex, zygos, caladan, snap, persephone, demikernel, dune, ghost, syrup, ccuserspace, userfaultfd, fuse, terra-incognita, uio, apple-driverkit, windows-umdf, pond.asplos23}, \name must also achieve competitive performance while moving the high-latency PCIe interconnect directly into the decision-making fast path.
We describe how we achieve this in \S\ref{sec:wave-implementation}.

\subsection{\name API}
\label{sec:wave-design:wave-api}

\name provides the API in \Cref{tab:wave-api} for host-SmartNIC communication.
The host uses the API to notify SmartNIC agents of state updates via messages and handle decision transactions.
The SmartNIC configures queues with the host, reads host messages via polling, and sends decisions, issuing an MSI-X vector to kick the host, as needed.
Similar to ghOSt~\cite{ghost}, these decisions are committed atomically in the host kernel using transactions.
For example, if an agent attempts to update page table entries for an application that simultaneously exits, the transaction will cleanly fail without corrupting host kernel state.
The \name API is general, facilitates batching, and supports a wide range of system software, as described in \S\ref{sec:wave-subsystems}.
Furthermore, \name permits each system software component to implement private extensions to the API, e.g., a memory manager reads access bits from host page table entries, and a file system queries for attached devices.

\textbf{Communication.} \name uses the lessons learned from prior work (\S\ref{sec:why-offload-the-os}) to implement low-latency communication across PCIe, which is essential for achieving performance competitive with that of on-host system software (\S\ref{sec:evaluation:thread-scheduling}).
Queues are backed by DMA or MMIO, and DMA queues can operate synchronously or asynchronously.
Multiple messages and transactions can be batched.

% Replace [table] with [table*] (both here and in the \end statement at the bottom) to make the table span the entire width of the page rather than just one column.
\begin{table}[t]
\fontsize{7.5pt}{10pt}\selectfont
\centering
\begin{tabular}{c|c}
\multicolumn{2}{c}{\textbf{\name Shared API Calls (both the host and SmartNIC can call these)}} \\
\hline
\multicolumn{2}{c}{\texttt{START\_WAVE\_AGENT()}} \\
\multicolumn{2}{c}{\texttt{KILL\_WAVE\_AGENT()}} \\
\multicolumn{2}{c}{} \\

\textbf{\name Host API}
&
\textbf{\name SmartNIC API}
\\
\hline
\multicolumn{2}{c}{\textbf{Queues}} \\
\hline
& \texttt{CREATE\_QUEUE()} \\
& \texttt{DESTROY\_QUEUE()} \\
& \texttt{ASSOC\_QUEUE\_WITH(agent, host core)} \\
& \texttt{SET\_QUEUE\_TYPE(mmio/dma async/dma sync)} \\
\hline
\multicolumn{2}{c}{\textbf{Messages}} \\
\hline
\texttt{SEND\_MESSAGES(q = queue)}
&
\texttt{POLL\_MESSAGES(q)}
\\
\hline
\multicolumn{2}{c}{\textbf{Transactions}} \\
\hline
\texttt{PREFETCH\_TXNS(q)} (\S\ref{sec:wave-implementation:hiding-agent-policy-loop-latency})
&
\texttt{TXN\_CREATE(q)}
\\
\texttt{POLL\_TXNS(q)}
&
\texttt{TXNS\_COMMIT(q, send/skip msi-x)}
\\
\hline
\multicolumn{2}{c}{\textbf{Transaction Outcomes}} \\
\hline
\texttt{SET\_TXNS\_OUTCOMES(q)}
&
\texttt{POLL\_TXNS\_OUTCOMES(q)}
\\
\hline
\multicolumn{2}{c}{\textbf{Custom System Software APIs Can Be Included (As Applicable)}} \\
\hline
\end{tabular}
\caption{Key \name API functions. We include certain parameters when they are not intuitive, but omit them otherwise to save space. Multiple messages and transactions can be batched, which is why some API function names are plural.}
\label{tab:wave-api}
\end{table}

\subsection{Security}
\label{sec:wave-design:security}
Agents are isolated in userspace on the SmartNIC CPU, which restricts their ability to access memory and send MSI-X's.
They are further isolated logically, i.e., they cannot read messages or stage decisions for resources they do not manage.
Each system software component has an on-host watchdog that kills its agent(s) when it detects they are malfunctioning. For example, the thread scheduler (\S\ref{sec:wave-subsystems}) watchdog terminates an agent that has not made a decision for >20 ms. 
\section{Offloading System Software with \name}
\label{sec:wave-subsystems}

We offload three preexisting pieces of system software with \name on Linux: thread scheduling~\cite{ghost}, memory management~\cite{sol}, and an RPC stack~\cite{stubby, grpc}.

\subsection{Kernel Thread Scheduling}
\label{sec:wave-subsystems:kernel-thread-scheduling}

The \textit{thread scheduler} places runnable threads onto idle cores and maintains the scheduling policy's run queues and data structures.
Application threads operate on timescales ranging from several $\mu$s to seconds, though schedulers must be responsive on the $\mu$s-scale to ensure critical threads, such as network RX, receive CPU time.
Policies include CFS~\cite{cfs}, Caladan~\cite{caladan}, Shinjuku~\cite{shinjuku}, and others~\cite{shenango, persephone, eevdf-1, eevdf-2}. 

To schedule threads, \name offloads policies built with ghOSt~\cite{ghost}, a general scheduling class in Linux that runs arbitrary policies in userspace \textit{agents}.
The agents receive thread state messages from the kernel (e.g., a thread blocked, woke up, etc.), make scheduling decisions, and commit those decisions to the kernel via a transaction API.
The kernel then context switches to the scheduled threads.
Communication between the kernel and userspace occurs via shared memory.

As shown in \Cref{fig:wave-detailed}, we move the ghOSt agents to the SmartNIC and keep the ghOSt kernel scheduling class on the host.
The communication patterns are the same as in ghOSt, and to further improve latency, the SmartNIC agents are always awake and polling for messages.

\textbf{Communication.}
The offloaded thread scheduler uses MMIO queues in both directions because scheduling decisions must be made in <1$\mu$s to avoid keeping host cores idle.
For example, a GET request for the RocksDB~\cite{rocksdb} key-value store requires 4$\mu$s of CPU time, so two 1$\mu$s PCIe transfers to schedule a thread degrade throughput by 50\%.
Low PCIe throughput of $\sim$13.5 KiB/s is required in both directions since the decisions are compact and generated only on thread events, such as when a thread blocks.

\textbf{Implementation.}
The \name thread scheduler integrates ghOSt with the Wave API in \S\ref{sec:wave-design:wave-api}, including \texttt{SEND\_MESSAGES()}, \texttt{PREFETCH\_TXNS()} (\S\ref{sec:wave-implementation:hiding-agent-policy-loop-latency}), \texttt{POLL\_TXNS()}, and \texttt{SET\_TXNS\_OUTCOMES()} on the host, and \texttt{POLL\_MESSAGES()}, \texttt{TXN\_CREATE()}, \texttt{TXNS\_COMMIT()}, and \texttt{POLL\_TXNS\_OUTCOMES()} on the SmartNIC.
To seamlessly support existing ghOSt policies, we extend the \name API with ghOSt-specific APIs ("Custom APIs" in \Cref{tab:wave-api}).

\textbf{Optimizations.}
The \name scheduler \textit{prestages} (\S\ref{sec:wave-implementation:hiding-agent-policy-loop-latency}) one decision per core so the host can \textit{prefetch} them when a thread blocks or yields.
The scheduler eagerly prestages decisions when the 
run queue length is sufficiently deep (e.g., linear in the number of cores), improving scheduling throughput by 32\% in \S\ref{sec:evaluation:thread-scheduling}.

\subsection{Memory Management}
\label{sec:wave-subsystems:memory-management}

The \textit{memory manager} decides how to map virtual pages to physical pages.
This includes when responding to allocation requests, swapping and compressing pages, and managing page table entries.
A memory manager's data structures scale linearly with the size of each process address space, requiring significant memory.
Policy algorithms, such as LRU, also require significant compute, so policy designers resort to approximations like the LRU CLOCK algorithm~\cite{clock}.
Other policies include SOL~\cite{sol}, AIFM~\cite{aifm}, and AutoNUMA~\cite{autonuma}.

\textbf{Design and Implementation.}
We offload memory management to the SmartNIC with \name and keep the in-kernel mechanism, such as page fault handlers, page table entries, and TLB shootdowns, on the host.
The host uses \texttt{SEND\_MESSAGES()} to send PTEs, which contain mappings and dirty/access bits, to the SmartNIC.
\name agents make page migration decisions and use \texttt{TXN\_CREATE()} and \texttt{TXNS\_COMMIT()} to send updated mappings to the host.
Upon receipt of the updated mappings, the host uses the \texttt{madvise()} syscall path in the kernel for migrations.

\textbf{Communication.}
Transferring PTEs to the SmartNIC and updated mappings back requires a high throughput of 1+ Gbps, so both queues use DMA.
Recall that page fault handlers and other fast-path mechanisms are kept on the host.
The memory manager itself operates out-of-band, so policy overhead of hundreds of milliseconds is acceptable.

\textbf{SOL Policy.}
Recent work proposes machine learning (ML)  policies to improve system efficiency across multiple metrics~\cite{lake, deeprm, decima, autophase, pensieve}, including DRAM consumption~\cite{sol}.
\textit{ML requires significant compute, so it is costly to deploy without an offload framework like \name.}
SOL~\cite{sol} is a recently proposed ML-based policy to classify hot (i.e., frequently accessed) memory pages into a "fast tier" (local DRAM) and cold pages into a "slow tier" (remote DRAM, non-volatile memory, or disk).

At startup, the SOL policy groups consecutive pages into batches for classification.
The policy scans page access bits and classifies batches using {\it Thompson Sampling with a Beta distribution prior}~\cite{sol, thompson-sampling}.
Each batch is scanned with a frequency ranging from once every 300ms to 9.6s, and batches are moved between memory tiers once per 38.4s epoch (i.e., 4x the slowest scanning frequency).
To reduce overhead, SOL determines the optimal frequency to scan each batch's access bits as each scan requires (1) flushing the TLB and (2) policy computation.
SOL's overhead is proportional to the address space size and other policy parameters.
We offload the SOL policy with \name and evaluate its performance in \S\ref{sec:evaluation:memory-management}.

\subsection{RPC Stack}
\label{sec:wave-subsystems:network-rpcs}

Recent research~\cite{dagger, flextoe, nanopu, erss, mind-the-gap, azure-accelerated-networking, lynx} and RDMA protocols~\cite{1rma,roce,networksupport} show clear value in offloading RPC stacks to SmartNICs.
RPC stacks process packets in a few $\mu$s~\cite{nsight} and require significant compute for protocol processing, serialization, compression, and security.
They have a large memory footprint due to large packet payloads and stream state.
Thus, offload saves host resources and enables acceleration.

We use \name to offload the Stubby RPC stack~\cite{stubby}, similar to gRPC~\cite{grpc}, including both a software-based packet-to-host-core steering policy and the data plane.
The vanilla on-host Stubby system uses hardware-based RSS~\cite{rss} on the NIC to steer packets to cores, the host Linux network stack for TCP processing, and CFS for scheduling.
After vanilla Stubby processes an incoming RPC, it executes an application-specified callback function to handle the request.
The application then executes an RPC callback to send an RPC response, again relying on the host Linux network stack.

We now describe how we offload both the packet steering policy and the data plane to the SmartNIC. As before, this offloaded software is encapsulated in a SmartNIC agent.

\textbf{Path of an RPC.}
An arriving RPC packet is steered to the ARM cores on the SmartNIC.
The SmartNIC uses its own Linux network stack for TCP processing.
The RPC is then passed to the RPC agent, which determines which host core to steer the RPC to.
The agent enqueues the RPC in an MMIO queue with \texttt{TXN\_CREATE()}
and sends it to the host core with \texttt{TXNS\_COMMIT()}.
We specify that \texttt{TXNS\_COMMIT()} (see \Cref{tab:wave-api}) does not send an MSI-X because the host will instead poll the queue to sustain high RPC throughput.

On the host, an RPC-enabled application links with a stub RPC library to make offload transparent.
The application polls the request via \texttt{POLL\_TXNS()}, handles it, and writes the RPC response to an MMIO queue via \texttt{SET\_TXNS\_OUTCOMES()}.
The polling agent picks up the response via \texttt{POLL\_TXNS\_OUTCOMES()} and sends it via the SmartNIC Linux network stack.

\textbf{MMIO for Communication.}
RPC stacks process RPCs in a few $\mu$s~\cite{nsight} before handing them to the application.
\S\ref{sec:evaluation:networking} uses small RPC payloads, so our RPC stack processes a moderate throughput of 10s of MiB/s of RPCs.
This combination of low latency and low throughput is what drove our decision to use MMIO for RPC host-SmartNIC communciation.
A hybrid approach of MMIO with DMA for large packet payloads, proposed by prior work~\cite{understanding-pcie-performance}, or just DMA alone, would be better for workloads with larger payloads.

\textbf{Queue Configuration.} The \name agent steers RPCs to specific host cores by stashing them in per-core SmartNIC-to-host queues.
There are also per-core host-to-SmartNIC queues for host cores to transfer RPC responses to the agent.

\textbf{Optimizations.} Related policies and systems like Receive-Side Scaling~\cite{rss}, IX~\cite{ix}, Shinjuku~\cite{shinjuku}, Shenango~\cite{shenango}, Demikernel~\cite{demikernel}, and others~\cite{grpc, dpdk, caladan} remove the RPC stack policy from the host kernel.
This gap between the kernel and network workload causes load imbalance.
The host wastes resources by either over-allocating CPU resources to handle traffic or dedicating host cores to making scheduling decisions after the SmartNIC transfers traffic to host memory~\cite{snap,shinjuku}.
In \S\ref{sec:evaluation:networking}, we show the benefits of co-locating the RPC stack with thread scheduling on the SmartNIC.
\section{Closing the Host-SmartNIC Latency Gap}
\label{sec:wave-implementation}

Offloading system software with \name places the slow PCIe interconnect between the host and the SmartNIC agents.
As described in \S\ref{sec:wave-subsystems}, each system software component has distinct communication requirements, covering a full spectrum of latency and throughput needs.
Fast communication over PCIe and efficient message-passing are extensively studied by prior work and are well understood.
Therefore, this section first discusses key insights from this work and how they influenced Wave’s communication design (\S\ref{sec:wave-implementation:prior-work}).
We then explore the communication mechanisms we considered (\S\ref{sec:wave-implementation:which-communication-mechanisms-did-we-explore}) before detailing our implementation of \name's host-SmartNIC queues (\S\ref{sec:wave-implementation:communication-solution}).
\name re-uses the PCIe DMA queue implementation from Floem~\cite{floem} and adds MMIO support.
For MMIO, we explain how using different page table entry types further reduces latency (\S\ref{sec:wave-implementation:communication-solution}).
Finally, we discuss how we hide the remaining latency to achieve fast system software decision-making (\S\ref{sec:wave-implementation:hiding-agent-policy-loop-latency}).

\subsection{Prior Work on Fast PCIe Communication}
\label{sec:wave-implementation:prior-work}

Prior work~\cite{understanding-pcie-performance, pio, characterizing-off-path-smartnic, azure-accelerated-networking, floem, mind-the-gap, e3, clicknp, memif} demonstrates that optimizing the slow PCIe communication path is \textit{critical} to making offload practical.
Without these optimizations, end-to-end performance with offload is worse than not offloading at all, even with the additional SmartNIC resources.

Neugebauer et al.~\cite{understanding-pcie-performance} propose a theoretical PCIe model and testbench for evaluating PCIe performance. They find that PCIe suffers from high latency (1000ns roundtrip) and that PCIe overheads degrade throughput by >20\%.
Optimizations are critical to improve performance, such as caching with DDIO ($\sim$15\% throughput and $\sim$70ns latency improvements), accounting for NUMA locality (10-20\% throughput difference),  using hugepages to avoid IOMMU overhead (up to 70\% throughput difference), descriptor batching, prefetching, disabling interrupts, and minimizing synchronization.

iPipe~\cite{ipipe} uses two unidirectional queues for host-SmartNIC communication and demonstrates a ~2-7x speedup when using asynchronous (i.e., non-blocking) DMA rather than synchronous and up to 8.7x speedup when batching transfers.
iPipe also uses lazy queue head synchronization to avoid PCIe roundtrips.
Floem~\cite{floem} also implements unidirectional queues with optimizations like asynchronous synchronization (9-15x speedup), pruning unused packet data (1.2-3.1x speedup), caching, batching, and hiding I/O.

\name leverages the insights from this prior work and builds on Floem's queue implementation for fast host-SmartNIC communication.
DMA is best for high-throughput transfer that is latency insensitive because several MMIO accesses are required to initiate DMA.
MMIO is best for low-latency, low-throughput communication.
\name similarly uses batching, caching, and prefetching; disables interrupts under heavy load; avoids unnecessary synchronization to reduce roundtrips; and accounts for NUMA locality.

\textbf{OS Contexts.} Microkernels~\cite{l4-microkernel-lessons, redleaf, l3-to-sel4, microkernel-goes-general, okl4, qnx, theseus, xpc, fuschia-zircon, harmonizing-microkernels, minix3, pikeos, sel4, hydra, improving-ipc-by-kernel-design, on-microkernel-construction, filesystem-semi-microkernel, mach}, multikernels~\cite{barrelfish, barrelfish-dc, barrelfish-intel-single-chip, barrelfish-cosh, barrelfish-msr, menzi-masters-thesis, barrelfish-technical-note-001, barrelfish-technical-note-005, nros}, exokernels~\cite{exokernel}, and userspace resource management systems~\cite{ix, shinjuku, shenango, reflex, zygos, caladan, snap, persephone, demikernel, dune, ghost, syrup, ccuserspace, userfaultfd, fuse, terra-incognita, uio, apple-driverkit, windows-umdf, pond.asplos23} optimize kernel-userspace communication.
These systems typically employ \textit{message-passing} over shared-memory queues.
\name uses message-passing and leverages their optimizations, including asynchronous communication, limited copying, batching, and cache-line alignment.

\subsection{Our SmartNIC's Communication Mechanisms}
\label{sec:wave-implementation:which-communication-mechanisms-did-we-explore}

Our SmartNIC supports DMA and MMIO over non-coherent PCIe.

\textbf{Direct Memory Access (DMA).}
The SmartNIC has a DMA engine that supports bidirectional memory transfer between host DRAM and the SmartNIC SoC DRAM.
When the host CPU is the producer, entries are written to host DRAM and DMA'd to the SmartNIC SoC DRAM.
Conversely, when the SmartNIC is the producer, the opposite occurs.

\textbf{Memory-Mapped I/O (MMIO).}
Like most other SmartNICs, the SmartNIC that we use exposes an MMIO interface to the host so that the host can read/write a subset of the SmartNIC memory, which the SmartNIC cores otherwise have fast, coherent access to via the SoC.
The MMIO interface is backed by a hardware function in the SmartNIC that operates without CPU intervention.
As shown in Table~\ref{tab:microbenchmarks}, the direct CPU overhead of a 64-bit host MMIO write is $\sim$50 ns and a read is $\sim$750 ns.
Writes have less CPU overhead than reads because they are not acknowledged, while reads must wait for the PCIe roundtrip to complete before the read values are available.
The host can use the MMIO region to access the SmartNIC DRAM, but the SmartNIC cores cannot access host memory without DMA.
The MMIO mechanism throughput and latency are bounded by the hardware capabilities of the PCIe interconnect and the SmartNIC.
Even so, the host MMIO overheads are not trivial and could still become bottlenecks for microsecond-scale workloads.
We reduce these overheads further in \S\ref{sec:wave-implementation:mmio-latency}.

\textbf{Coherent Interconnects.}
New coherent interconnects such as CXL~\cite{cxl}, UPI~\cite{upi}, and NVLink~\cite{nvlink} provide a shared coherent memory address space between the SmartNIC and host.
Once such interconnects are widely used, they will accelerate host MMIO accesses.
SmartNIC SoC memory will be cacheable on the host, so the host can hide read latency by prefetching and reusing MMIO reads.
MMIO writes will also be faster since the host can store a batch of writes in the host cache hierarchy and flush the batch to PCIe.
However, DMA will remain preferred for high-throughput, latency-insensitive workloads.
MMIO with CXL requires CPU coordination for memory transfer, while DMA offloads this coordination to the DMA engine, freeing up CPU cycles.
And even though coherent interconnects improve low-latency MMIO, the \name optimizations in \S\ref{sec:wave-implementation:mmio-latency} and \S\ref{sec:wave-implementation:hiding-agent-policy-loop-latency} remain necessary to make performant offload practical.

\subsection{Host-SmartNIC Communication in \name}
\label{sec:wave-implementation:communication-solution}

Like microkernels and multikernels, \name uses message-passing over shared-memory queues for host-SmartNIC communication.
It re-uses the Floem~\cite{floem} DMA unidirectional queue and adds support for MMIO to achieve the best of both worlds: low-latency and high-throughput communication.
We further optimize the MMIO queue to reduce latency.

\textbf{DMA.} 
The producer writes one or more entries to the Floem queue and initiates a DMA transaction to send the entries to the consumer.
The producer can wait synchronously until the transaction completes or it can check for completion \textit{asynchronously}.
Messages can be \textit{batched}, and they are written to the \textit{local NUMA node} of the recipient.
To ensure the consumer does not read inconsistent entries while the producer is still writing, the producer sets a flag in each entry \textit{after} it finishes writing the entry.
If the consumer sees the per-entry flag set, the entry is valid to read.

\textbf{MMIO.}
\name MMIO message queues are identical to Floem DMA queues, with the same data layout and flag synchronization scheme.
As only the SmartNIC exposes its DRAM via MMIO, the queues are always backed by SmartNIC DRAM, regardless of which side is the producer or consumer.
Thus, the host accesses these queues via MMIO while the SmartNIC agents access them via local, coherent memory.
MMIO queues also leverage optimizations like batching (\S\ref{sec:wave-implementation:mmio-latency}), caching (\S\ref{sec:wave-implementation:mmio-latency-caching}) and prefetching (\S\ref{sec:wave-implementation:hiding-agent-policy-loop-latency}).
\vspace{1.0em}
\subsubsection{Reducing MMIO Latency}
\label{sec:wave-implementation:mmio-latency}

\begin{figure}[t]
\centering
\includegraphics[width=0.8\columnwidth,keepaspectratio]{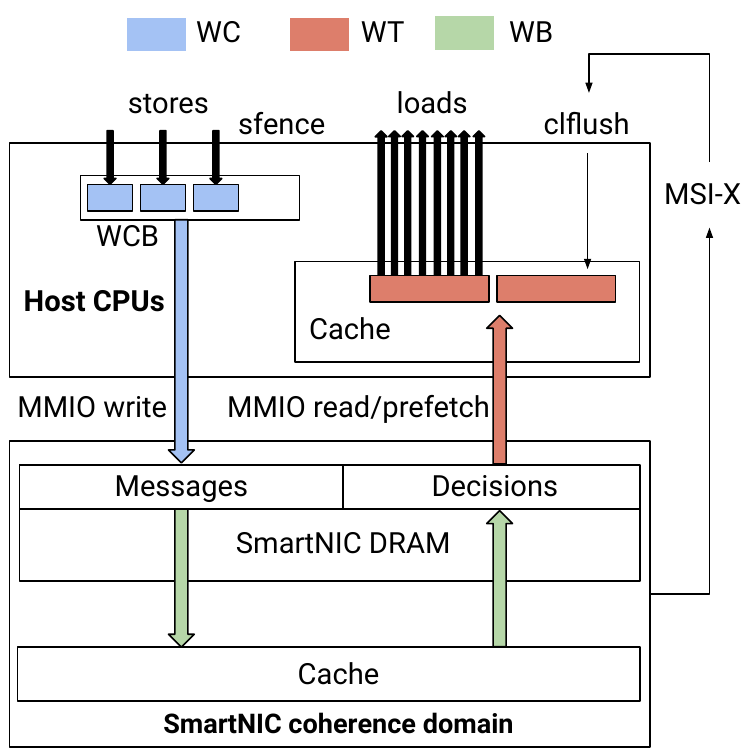}
\caption{\name improves performance by using caches and multiple page table entry types.}
\Description[Wave cache and PTE configuration.]{Wave improves performance by using caches and multiple page table entry types.}
\label{fig:ptes}
\end{figure}
% https://docs.google.com/drawings/d/1kc58abRfuw2a8K3Kpyy_7qTzSm1VgW8qPCz8cVp49eM/edit

MMIO is in the critical path of system software that requires very low (<1 $\mu$s) latency, such as thread scheduling (\Cref{fig:evaluation:rocksdb}) and an RPC stack (\Cref{fig:evaluation:rpc}).
Host MMIO reads, in particular, are very expensive (750ns) due to the PCIe roundtrip.
To further reduce MMIO overheads, \name leverages the \textit{caching} and \textit{prefetching} optimizations proposed by prior work.
To do so, it is important to set page table entry (PTE) types to cache MMIO contents.
\Cref{fig:ptes} shows how \name sets PTE types and uses caches.
\name uses these PTE types:

\textbf{Write-Back (WB).} The CPU caches and implements coherence for this type of memory (e.g., local DRAM). This is the default mapping for userspace memory.

\textbf{Write-Combining (WC).} The CPU does not cache reads or implement coherence, instead loading directly from memory.
Stores go to a write-combining buffer, so they complete much faster than going to memory.
The buffer drains to memory periodically and is flushed explicitly with \texttt{sfence}.

\textbf{Write-Through (WT).} The CPU stores directly to memory. However, loads are cached, so subsequent loads from the same cache lines are fast. 

\textit{SmartNIC CPU PTEs.} MMIO queues are backed by SmartNIC DRAM. The agents have local, coherent access to this memory, which is mapped into their address space with WB.

\textit{Host CPU PTEs.} There is non-coherent PCIe between the host CPU and SmartNIC memory, so the host cannot map the MMIO region with WB. However, the host can improve write performance with WC instead of having no caching at all (i.e., uncacheable PTEs).
The host can now enqueue a message \textit{batch} before the write-combining buffer is flushed.

\subsubsection{Caching MMIO Content}
\label{sec:wave-implementation:mmio-latency-caching}
A 64-bit host MMIO read takes $\sim$750 ns, so it is expensive for the host to read decisions, especially those that span multiple words.
To reduce overhead, we mark the MMIO queues as WT.
On each 64-\textit{bit} read to an MMIO address, the host CPU transfers the entire 64-\textit{byte} cache line into its cache, with the same $\sim$750 ns overhead.
Subsequent reads to the same cache line hit the cache, making it cheap to read a batch of decisions.

With this approach, however, stale decisions linger in the host CPU cache, so
\name implements a software-based coherence protocol.
When an agent stashes a new decision, the host flushes the stale data from the host CPU cache with \texttt{clflush}.
For example, the host flushes upon receiving an MSI-X from the SmartNIC as it knows a new decision is available, so the currently cached decision is stale.

\subsubsection{Conclusion}
Writing messages is cheap with WC.
Reading decisions is cheap with WT as read overhead is amortized across a cache line and subsequent reads hit the cache.
We next hide the remaining read latency entirely.

\subsection{Hiding the Remaining Latency}
\label{sec:wave-implementation:hiding-agent-policy-loop-latency}

Despite our optimizations, systems offloaded with \name still place PCIe in the critical path, which causes higher decision latency than on-host solutions.
\name lets these systems hide the remaining latency through prestaging and prefetching.

\textbf{Prestaging Decisions.}
When the host requires a decision (\Cref{fig:wave-detailed}), it experiences a long wait of several $\mu$s before the SmartNIC agent receives a message and sends a decision back.
During this wait, the host CPU cannot execute the workload, leading to low host utilization.
A \name agent can make decisions ahead of time--before the host requests a decision--and \textit{prestage} those decisions in SmartNIC DRAM.
When the host requires a decision, the host immediately checks the MMIO region for a decision and acts on it if present.
The \name agent is not invoked on the critical path, significantly reducing decision latency.
This approach leads to a substantial improvement in workload performance (32\% in \S\ref{sec:evaluation:thread-scheduling}) and decouples host utilization from decision latency.

\textit{When is prestaging possible?} Prestaging is possible when the agent can make decisions before asked.
A memory management agent can detect a strided host access, preallocate physical frames that will soon be needed, and then prestage new page table entries.
A thread scheduler with a deep run queue can prestage a runnable thread on each core.

\textbf{Prefetching MMIO Decisions from the Agent.}
When the host needs a decision, it updates kernel state and sends a message to the \name agent on the SmartNIC; it then reads a prestaged decision.
The decision queue MMIO is marked as WT, which supports prefetching.
Rather than issue a blocking load for the prestaged decision right when the decision is needed, the host can now issue a memory prefetch for the decision \textit{before} updating the kernel state and sending a message.
This work takes about 1$\mu$s, which is sufficient to completely hide the MMIO read latency, so the subsequent host load for the decision will hit the host cache.
Thus, when decisions are prestaged, prefetching a decision reduces the cost of MMIO reads to zero.
\S\ref{sec:evaluation:thread-scheduling} shows that with these optimizations, \name closes the host-SmartNIC latency gap.
\section{Lessons Learned from \name Agents}
\label{sec:wave-subsystems:lessons}

We now summarize lessons learned from our experience offloading preexisting userspace system software with \name.

\textbf{Minimize or Hide PCIe Overheads.}
The PCIe interconnect is in the host-SmartNIC critical path.
System software must account for PCIe overheads to remain performant, akin to how an OS must consider cross-NUMA transfers in a multi-socket machine.
It should leverage insights from prior work on fast communication between OS contexts and across PCIe.
These insights include choosing the right transport mechanism (setting DMA or MMIO via \texttt{SET\_QUEUE\_TYPE()}), minimizing state sharing~\cite{floem}, batching (e.g., multiple txns in \texttt{TXNS\_COMMIT()}), asynchronous communication, caching, prefetching (\texttt{PREFETCH\_TXNS()}), and NUMA locality.

\textbf{Keep Agents Modular.}
Management, engineering, and debugging complexity is reduced when agents encapsulate policy state as they seamlessly operate across SmartNICs and operating systems, and in host userspace when a SmartNIC is unavailable.
It is also easier to experiment with different system software components and offload techniques~\cite{floem}.

\textbf{Keep Fault Recovery Simple.}
An agent may crash or be killed in preparation for an upgrade.
An operator may wish to restart the agent or fall back to vanilla on-host system software.
In both cases, recovery is easier when the host kernel is the \textit{source of truth} for non-policy state, such as page table entries for a memory manager or thread TIDs for a scheduler.
Both a restarted agent and vanilla on-host system software pull this information upon launch and continue running.
Complicated checkpointing or state recovery increase complexity and create  bugs that may require a machine reset.

\textbf{Focus on Host Partitions and Agent Scalability.}
Datacenter machines have hundreds of CPU cores, accelerators, TiBs of DRAM and flash, and 100+ Gbps NIC bandwidths.
Multiple applications/tenants share these resources but want different policies.
Developers should partition host resources into logical units, each with their own agent and policy, following the proven approach of ghOSt \textit{enclaves}.
The scheduling agent in \S\ref{sec:evaluation:thread-scheduling} operates per CCX and the memory agent in \S\ref{sec:evaluation:memory-management} manages a single address space.
Developers should also parallelize an agent with threads.
Each memory agent thread in \S\ref{sec:evaluation:memory-management} manages an address space chunk.
 \section{Evaluation}
\label{sec:evaluation}

\name is implemented on the 4.15 Linux kernel on the host and the 5.10 kernel on the SmartNIC.\footnote{We use these versions of Linux on our production machines. \name and the offloaded system software also work with the newest versions of Linux.}
We use the Intel Mount Evans SmartNIC~\cite{intel-mount-evans}, which has an ARM Neoverse N1~\cite{neoverse} CPU @ 3.0 GHz with 16 physical cores and a network function with 200Gbps of throughput.
The host has an AMD Zen3 CPU @ 2.45 GHz with 2 sockets, 64 physical cores per socket, and 2 hyperthreads per physical core.
The host CPU frequency ranges from 2.45 GHz to up to 3.5 GHz with turbo boost.
The CPU contains core complexes (i.e., AMD CCX's) with 8 physical cores and a private L3 cache per CCX.
Unless otherwise noted, we run workloads on the first hyperthread of each host physical core and leave the second sibling idle.

We present microbenchmarks that show \name has low communication overheads that make offload practical.
We then offload preexisting system software for thread scheduling, an RPC stack, and memory management to the SmartNIC and provide an apples-to-apples comparison to on-host software.
Due to our optimizations, end-to-end performance is only slightly worse with \name despite PCIe overheads.
When applications leverage newly freed host resources due to the offloaded software, they outperform on-host deployments.

\subsection{Microbenchmarks}
\label{sec:evaluation:microbenchmarks}

\textbf{Results of MMIO and MSI-X Evaluation.} In \Cref{tab:microbenchmarks}, host MMIO reads are 750ns due to the PCIe roundtrip and writes are 50ns as they are not acknowledged.
\name uses SmartNIC memory for communication and, thus, NIC-side latencies are those of a regular memory access.
MSI-X overheads are comparable to interprocessor interrupts, though end-to-end latency is higher due to the one-way PCIe trip.

\textbf{Results of Scheduling Evaluation.} \Cref{tab:microbenchmarks-sched} contextualizes these overheads by comparing a scheduler offloaded with \name with on-host ghOSt for a FIFO thread scheduling policy.
With no optimization, \name's overheads are several $\mu$s higher than ghOSt's, though optimizations (\S\ref{sec:wave-implementation}) bring \name quite close.

\textbf{Conclusion.} \name has cheap communication and notification mechanisms that make offload practical.
\name's microbenchmarks will never match on-host due to the inherent cost of offload across PCIe, though they come close.
\begin{table}[t]
\centering
\resizebox{0.8\columnwidth}{!}{
\footnotesize
\centering \begin{tabular}{lr}
\hline
\textbf{MMIO}
&
\\
1. Host MMIO 64-bit Read (Uncacheable)
&
750 ns
\\
2. Host MMIO 64-bit Write (Uncacheable)
&
50 ns
\\
\hline
\textbf{MSI-X}
&
\\
3. MSI-X Send (Register Write)
&
70 ns
\\
4. MSI-X Send (Ioctl + Register Write)
&
340 ns
\\
5. MSI-X Receive
&
350 ns
\\
6. MSI-X End-to-End
&
1,600 ns
\\
\hline
\\
\end{tabular}
}
\vspace{-1.0em}
\caption{Hardware microbenchmarks. Results are rounded to 1-2 leading digits. MSI-X End-to-End includes PCIe latency.}
\label{tab:microbenchmarks}
\end{table}

\begin{table}[t]
\centering
\resizebox{\columnwidth}{!}{
\begin{tabular}{lr}
\hline
\textbf{Offloaded Kernel Thread Scheduler with \name}
&
\\
1. Open a Decision in Agent \& Send MSI-X
&
\\
\hspace{0.5cm}Baseline (\S\ref{sec:wave-implementation:communication-solution})
&
1,013 ns
\\
\hspace{0.5cm}with WB PTEs on SmartNIC (\S\ref{sec:wave-implementation:mmio-latency})
&
426 ns
\\
2. Context Switch Overhead on Host
&
\\
\hspace{0.5cm}Baseline (\S\ref{sec:wave-implementation:communication-solution})
&
% Across 5 runs of `./sync --type block`, I saw medians of 13.53us, 13.44us, 13.351us, 13.45us, and 13.31us.
13,310-13,530 ns
\\
\hspace{0.5cm}with WB PTEs on SmartNIC (\S\ref{sec:wave-implementation:mmio-latency})
&
% Across 5 runs of `./sync --type block`, I saw medians of 9.99us, 10.009us, 9.99us, 10.16us, and 9.94us.
9,940-10,160 ns
\\
\hspace{0.5cm}and with WC/WT PTEs on Host (\S\ref{sec:wave-implementation:mmio-latency})
&
% Across 5 runs of `./sync --type block`, I saw medians of 6.84us, 6.21us, 6.91us, 6.86us, and 6.1us.
6,100-6,910 ns
\\
\hspace{0.5cm}and with Pre-Staging \& Prefetching (\S\ref{sec:wave-implementation:hiding-agent-policy-loop-latency})
&
% Across 5 runs of `./sync --type block`, I saw medians of 3.32us, 4.02us, 4.04us, 3.909us, and 3.32us.
3,320-4,040 ns
\\
\hline
\textbf{On-Host ghOSt Scheduler} & \\
3. Open a Decision in Agent \& Send Interrupt
&
770 ns
\\
4. Context Switch Overhead on Host
&
\\
\hspace{0.5cm}Baseline
&
% Across 5 runs of `./sync --type block`:
% On CPU 8, I saw medians of 4.4us, 4.99us, 4.38us, 4.54us, and 4.41us.
% On CPU 9, I saw medians of 4.35us, 5.12us, 4.38us, 4.48us, and 4.391us.
4,380-4,990 ns
\\
\hspace{0.5cm}with Pre-Staging (\S\ref{sec:wave-implementation:hiding-agent-policy-loop-latency})
&
% Across 5 runs of `./sync --type block`:
% On CPU 8, I saw medians of 2.37us, 2.35us, 3.21us, 2.36us, and 2.36us.
% On CPU 9, I saw medians of 2.4us, 2.36us, 3.26us, 2.37us, and 2.36us.
2,350-3,260 ns
\\
\hline
\\
\end{tabular}
}
\vspace{-1.0em}
\caption{Scheduling microbenchmarks. We ran context-switch benchmarks five times and report the range of medians. "Prestaging" has more variability as prestages may fail.}
\label{tab:microbenchmarks-sched}
\end{table}

\subsection{Thread Scheduler}
\label{sec:evaluation:thread-scheduling}

We offload three preexisting ghOSt policies with \name and compare them with their on-host implementations in ghOSt to demonstrate the overheads of offload.
\subsubsection{Ported ghOSt Policies}
We first port a \textbf{run-to-completion FIFO policy} and use it to schedule RocksDB \cite{rocksdb}, an in-memory key-value store with $\mu$s-scale requests.
We chose this policy because it requires little compute but interacts extensively with the workload, stressing \name's API and PCIe queues and making the cost of offload clear.

We next port a \textbf{Shinjuku~\cite{shinjuku} policy} and use it to schedule RocksDB.
We chose Shinjuku because it also maintains a FIFO queue but preempts threads that exceed a time slice, making the overhead of MSI-X clear.

Last, we port \textbf{our custom kernel scheduling policy for GCE, our production virtual machine service}~\cite{gce}, to \name.
vCPUs in our VM service run for several milliseconds continuously before requiring scheduler intervention.
This policy shows that \name's API is sufficiently expressive to support production policies and that \name suffers negligible loss of performance when scheduling ms-scale workloads.

For all policies, we perform an apples-to-apples comparison between the policy on the SmartNIC and the policy on the host.
When the policy is on the SmartNIC, we do not let the workload use the newly freed host resources that the policy would have consumed if it were on the host.
This clarifies the cost of offload.
Subsequently, we let the workload use these newly freed resources to show the benefit of offload.
In all cases, both offloaded and on-host policies either consistently use or do not use the prestaging optimization together.
Offloaded policies alone may use the \name-specific optimizations of WC/WT page table entries and WT prefetching.

\subsubsection{FIFO Evaluation}
We compare a FIFO policy in both \name and on-host ghOSt and drive RocksDB with 10 $\mu$s GET requests.
In both cases, one agent thread is pinned to a core and schedules RocksDB.
Both prestage threads (\S\ref{sec:wave-implementation:hiding-agent-policy-loop-latency}).

\textbf{Comparison Scenarios.} We compare three scenarios: \textit{On-Host:} On-host ghOSt with 16 host cores. One runs a ghOSt agent and the other 15 run worker threads.
\textit{\name-15:} \name with 15 host cores, all of which run worker threads, the same number as On-Host.
\textit{\name-16:} \name with 16 host cores, all of which run worker threads. \name frees a host core that on-host ghOSt dedicates to an agent, so this core runs workers.

\textbf{Results of Apples-to-Apples Comparison.} In Fig. \ref{fig:get-10}, Wave-15's tail latency is 3$\mu$s higher than On-Host and saturates 1.1\% lower due to PCIe overhead.
\name-16 saturates 4.6\% higher than On-Host due to the extra host core freed.

\begin{figure}
\newcommand\fighalfpagewidth{0.48\textwidth}
\newcommand\figthirdpagewidth{0.29\textwidth}
\newcommand\figspacing{2mm}
\newcommand\figwidth{6.5cm}
\newcommand\figheight{4cm}

\begin{subfigure}[t]{0.6\linewidth}
\resizebox{\linewidth}{!}{\begin{tikzpicture}
\begin{axis}[
    title={},
    xlabel={RocksDB ~Throughput (x1K req/s)},
    ylabel style={align=center}, ylabel={99\% Latency ($\mu$s)},
    ylabel near ticks,
    xmin=0, xmax=910000,
    ymin=0, ymax=250,
    scaled x ticks=base 10:-3,
    hide scale,
    ymajorgrids=true,
    grid style=dashed,
    width = \figwidth,
    height = \figheight,
    legend style={at={(0.4,0.9)},anchor=north},
    every axis plot/.append style={very thick}
]

\addplot[
    color=green,
    mark=square,
    ]
    coordinates {
(49989,40)(99852,37)(149994,37)(200089,38)(249965,37)(300047,37)(350077,38)(400044,39)(450003,39)(499865,40)(550096,41)(600269,42)(649981,43)(700023,46)(750038,51)(800014,66)(819995,86)(820853,80)(822031,86)(823058,84)(823790,90)(824834,86)(825948,90)(826758,92)(827889,93)(828899,96)(829926,94)(830884,107)(831976,104)(832840,96)(833874,106)(834811,114)(836003,212)(836960,116)(838265,106)(839164,127)(840346,113)(840653,154)(842040,134)(842971,172)(844166,144)(845154,135)(846094,163)(846945,530)
    };
\addlegendentry{\small \name, 15 CPUs}

\addplot[
    color=red,
    mark=o,
    ]
    coordinates {
(50077,33)(99964,34)(150025,34)(200055,34)(250104,36)(299907,35)(349898,35)(400140,36)(450082,37)(499762,42)(550142,39)(600149,40)(649511,42)(699765,44)(749798,49)(799915,64)(849921,167)(850843,128)(851863,134)(852938,195)(854006,131)(855081,187)(855829,285)
    };
\addlegendentry{\small On-Host, 16 CPUs}

\addplot[
    color=blue,
    mark=triangle,
    ]
    coordinates {
(49954,36)(99830,37)(149945,37)(200040,38)(250101,37)(300139,37)(349853,38)(400053,38)(449936,39)(499814,40)(550179,41)(599942,42)(649966,42)(700050,44)(750295,46)(800120,51)(849951,63)(879863,89)(880839,98)(882006,88)(882831,89)(884062,91)(885283,93)(886144,126)(886737,103)(888297,104)(889197,107)(889906,111)(890955,117)(892041,228)(893084,123)(893943,113)(894724,162)(896099,3605)
    };
\addlegendentry{\small \name, 16 CPUs}

\end{axis}
\end{tikzpicture}}
\caption{FIFO scheduling policy for 10$\mu$s GET requests.
}
\label{fig:get-10}
\vspace{0.3cm}
\end{subfigure}

\begin{subfigure}[t]{0.6\linewidth}
\resizebox{\linewidth}{!}{\begin{tikzpicture}
\begin{axis}[
    title={},
    xlabel={RocksDB ~Throughput (x1K req/s)},
    ylabel style={align=center}, ylabel={99\% Latency ($\mu$s)},
    ylabel near ticks,
    xmin=0, xmax=175000,
    ymin=0, ymax=250,
    scaled x ticks=base 10:-3,
    hide scale,
    ymajorgrids=true,
    grid style=dashed,
    width = \figwidth,
    height = \figheight,
    legend style={at={(0.307,1.0)},anchor=north},
    every axis plot/.append style={very thick}
]

\addplot[
    color=green,
    mark=square,
    ]
    coordinates {
(10017,55)(19988,58)(29993,60)(39987,62)(49987,63)(59964,63)(69940,64)(79963,64)(90046,64)(99922,65)(110007,65)(119943,65)(130021,66)(140062,67)(140973,68)(141981,68)(142901,70)(143961,73)(144958,72)(145930,491)
    };
\addlegendentry{Offload, 15 CPUs}

\addplot[
    color=red,
    mark=o,
    ]
    coordinates {
(10012,48)(19994,54)(29943,53)(39993,55)(50044,56)(59985,59)(70058,57)(79988,61)(90039,59)(100055,61)(109992,60)(120055,61)(130079,62)(139961,63)(141090,63)(142029,63)(143144,64)(144013,64)(145070,64)(146108,70)(146957,66)(148058,64)(148934,122)(150002,156)(151039,144)(152126,81)(153047,138)(153859,101)(155010,216)(156035,101)(156998,479)
    };
\addlegendentry{On-Host, 16 CPUs}

\addplot[
    color=blue,
    mark=triangle,
    ]
    coordinates {
(10005,55)(20010,58)(30015,60)(39970,62)(50008,63)(59916,63)(69979,64)(79892,64)(90036,65)(99971,65)(109959,65)(119992,65)(130057,66)(139939,66)(141064,66)(142020,66)(142925,67)(144024,67)(144942,67)(146008,67)(147065,68)(147887,67)(148908,69)(149993,68)(151159,69)(151949,70)(152999,69)(153880,69)(155115,70)(156118,136)(156885,124)(157892,103)(158929,95)(160094,367)
    };
\addlegendentry{Offload, 16 CPUs}

% Hide the legend
\legend{}

\end{axis}
\end{tikzpicture}}
\caption{Shinjuku for 99.5\% 10$\mu$s GET and 0.5\% 10ms RANGE.
}
\label{fig:shinjuku}
\end{subfigure}

\vspace{-0.8em}
\caption{\name implements $\mu$s-scale preemptive scheduling and is competitive with on-host scheduling.}
\Description[Scheduling performance in Wave.]{Wave implements $\mu$s-scale preemptive scheduling and is competitive with on-host scheduling.}
\label{fig:evaluation:rocksdb}
\end{figure}
\textbf{Results of Applying Optimizations (\S\ref{sec:wave-implementation}). }
We repeat \name-16 without optimizations and successively add them.
\vspace{-0.8em}
\begin{table}[H]
\centering
\resizebox{0.86\columnwidth}{!}{
\begin{tabular}{|p{4.5cm}|p{2.5cm}|}
\hline
&
\parbox{2.5cm}{
\textbf{Saturation Tput}
}
\\
\hline
Baseline (No Optimizations, \S\ref{sec:wave-implementation})
&
258,000
\\
\hline
+ SmartNIC WB PTEs (\S\ref{sec:wave-implementation:mmio-latency})
&
520,000 (+102\%)
\\
\hline
+ Host WC/WT PTEs (\S\ref{sec:wave-implementation:mmio-latency})
&
680,000 (+31\%)
\\
\hline
\parbox{4.5cm}{
\vspace{0.059cm}
+ Prestage and Prefetch (\S\ref{sec:wave-implementation:hiding-agent-policy-loop-latency})
\vspace{0.07cm}
}
&
895,000 (+32\%)
\\
\hline
\end{tabular}
}
\label{tab:wave-optimizations}
\end{table}
\vspace{-1.0em}
Optimizing PCIe communication with WC/WT page table entries still falls far short of Wave-16's saturation throughput in \Cref{fig:get-10}.
When latency is \textit{hidden} with prestaging and prefetching, throughput further increases by 32\%. Wave-16 now saturates 4.6\% higher than On-Host, which is lower than an expected improvement of (16 / 15) - 1 = 6.7\% from an extra host core due to PCIe overhead.

\subsubsection{Shinjuku Scheduling Policy Evaluation}
Shinjuku~\cite{shinjuku} is a round-robin policy with time-based preemption.
We use this policy to show that MSI-X interrupts are a reasonable alternative to inter-processor interrupts.

\textbf{Results of MSI-X Evaluation. }Shinjuku preempts requests that exceed a time slice so short requests do not suffer inflated latency when stuck behind long requests.
The load generator produces 99.5\% of requests as 10$\mu$s GETs and 0.5\% as 10ms RANGE queries.
The preemption time slice is 30$\mu$s.
Both \name and on-host ghOSt prestage.
However, prefetching in \name (\S\ref{sec:wave-implementation:hiding-agent-policy-loop-latency}) is ineffective when a preemption occurs as the host immediately reads the next decision upon receipt of the MSI-X, so it cannot overlap the PCIe read with work.

\textbf{Results of Apples-to-Apples Comparison.} In Fig.~\ref{fig:shinjuku}, Wave-15 has tail latency 5$\mu$s higher than On-Host and saturates 7.6\% lower due to PCIe overhead.
\name-16 leverages the extra host core to saturate 1.9\% higher than On-Host, but this falls short of the expected 6.7\% improvement mentioned.
Wave-15 and Wave-16 both perform worse relative to On-Host than with FIFO due to lack of prefetch on preemption.

\subsubsection{Virtual Machine Scheduling Policy Evaluation}
\label{sec:evaluation:new-opportunities:reduce-turbo-interference}
Virtual machines run on top of a hypervisor in the host kernel, with their vCPUs scheduled by the host kernel's thread scheduler every few milliseconds.
Even when a vCPU is idle and there are no other runnable tasks on a physical core, the host scheduler still receives timer ticks and wakes up before immediately yielding back to the guest.
This unnecessary \textit{interference} prevents the physical core from entering deep sleep states, \textit{limiting the turbo boost potential of other cores}.

\textbf{Results of Offloading Custom Kernel Scheduling.} A prime candidate for offload is our custom kernel scheduling policy, designed specifically for GCE~\cite{gce}, our production virtual machine service.
This policy, inspired by Tableau~\cite{tableau}, prioritizes fair CPU sharing among all vCPUs while imposing an upper bound on tail latency. 
Under this policy, vCPUs run for a time quantum ranging from 5-10 ms but can be preempted at 1-ms granularity.
This fine-grained control ensures fairness as vCPUs may consume varying amounts of CPU time within their assigned quantum.

Each host core independently schedules its tasks, so our production machines deliver timer ticks to every core once per millisecond.
This high tick frequency prevents idle vCPUs from entering deeper power-saving modes, constraining the turbo boost capabilities of other cores.
If the on-host scheduler used a single polling scheduling instance instead, we could disable timer ticks.
However, ghOSt~\cite{ghost} shows that this model has scalability limits when co-located with the workload it schedules~\cite{ghost} due to hyperthread interference and coherence traffic.
We also could not provide the instance's core to cloud clients or provide machine-sized VMs, which reduce noise and administrative overheads and are an original motivation for SmartNICs like the Nitro~\cite{nitro}.

\textbf{Evaluation Background.}
We compare this VM scheduling policy in \name and on-host ghOSt.
As VMs are scheduled at ms-granularity, neither policy uses prestaging, and the \name policy also does not use prefetching.
In both cases, two VMs are scheduled, each with 128 vCPUs.
The VMs are multiplexed across 128 logical cores (64 physical cores) in a single socket. 
We limit our focus to one socket because turbo is constrained on a per-socket basis, i.e., one socket's activity does not impact the other sockets' clock frequency.
In each VM, we run the \texttt{busy\_loop} utility, which consumes cycles with arithmetic operations and system calls.
We use this utility internally to characterize compute performance and generate turbo frequency curves for new hardware.

\begin{figure}[t]
\newcommand\fighalfpagewidth{0.48\textwidth}
\newcommand\figthirdpagewidth{0.29\textwidth}
\newcommand\figspacing{2mm}
\newcommand\figwidth{6.5cm}
\newcommand\figbusyloopheight{4.0cm}
\newcommand\figimprovementheight{4.0cm}

\begin{subfigure}[t]{0.6\linewidth}
\resizebox{\linewidth}{!}{\begin{tikzpicture}
\begin{axis}[
    title={},
    xlabel={\# vCPUs Running \texttt{busy\_loop}},
    ylabel style={align=center},
    ylabel={Avg Per-vCPU\\Work ($10^{-8}$)},
    xmin=0, xmax=128,
    ymin=0, ymax=700000000,
    xtick={0, 16, 32, 48, 64, 80, 96, 112, 128},
    hide scale,
    ymajorgrids=true,
    grid style=dashed,
    width = \figwidth,
    height = \figbusyloopheight,
    legend style={at={(0.383,0.417)},anchor=north},
    every axis plot/.append style={very thick}
]

\addplot[
    color=red,
    mark=*,
    mark size=2.0pt,
    mark repeat=3,
    ]
    coordinates {
(1,633201950.5)(2,634292141.5)(3,634609991.2)(4,633262079.6)(5,633446399.5)(6,633820063.9)(7,633089415.6)(8,632956149.3)(9,634912731.1)(10,632910872)(11,633122175)(12,633568515)(13,632112691.5)(14,633395000.6)(15,634076047.6)(16,632745778.7)(17,632167850.1)(18,634268676.1)(19,630929351.4)(20,631984602.6)(21,633889360.4)(22,632699393.5)(23,631442265.9)(24,633733228.7)(25,631672562.9)(26,629617824.2)(27,633188369.7)(28,631899750)(29,628950697.8)(30,630009506)(31,623480096.9)(32,602823024.2)(33,594830642.1)(34,595016783)(35,598892740.2)(36,595198343)(37,594515934.3)(38,598580730.1)(39,596347982.3)(40,592998841.9)(41,596739869.7)(42,597382672)(43,593419533.1)(44,596350856.4)(45,596826630.3)(46,593038084.4)(47,592392501.8)(48,586862653)(49,579733475.1)(50,577733217)(51,575434806.2)(52,569798611.4)(53,567236862.6)(54,564510567.9)(55,559517341)(56,556304190.4)(57,554033168.8)(58,550254091.5)(59,546843702.3)(60,543888648.4)(61,540303794.6)(62,536918921.9)(63,533532076)(64,530551940.6)(65,522675415.6)(66,515059944.8)(67,508167288)(68,501516639.3)(69,494741934.5)(70,488287838.9)(71,481759386.2)(72,475535896.6)(73,469424402.5)(74,463791527.5)(75,458103761.1)(76,452624698.3)(77,447297966.4)(78,441727431.3)(79,436610890)(80,431566892.6)(81,426731012.8)(82,422076931)(83,417392956.3)(84,412882502.7)(85,408319703.6)(86,403756333.5)(87,399619668.9)(88,395485894.7)(89,391615198.1)(90,387752958.6)(91,383757228.3)(92,379845124)(93,376102051.8)(94,372376123.2)(95,368860662.7)(96,365488974)(97,362112630.8)(98,358767134.8)(99,355456943.2)(100,352229575.4)(101,349015468.2)(102,345865571.3)(103,342984602.5)(104,340200072.6)(105,337414276.9)(106,334350643.1)(107,331560922)(108,328549944.4)(109,325892820.8)(110,323236569.3)(111,320794232.4)(112,318220340.2)(113,315721671.7)(114,313203692.1)(115,310728147.2)(116,308216604.1)(117,305923845.6)(118,303582726.1)(119,301269663.4)(120,299046237.1)(121,296742146.5)(122,294591015.5)(123,292321857)(124,289962987.7)(125,287846545)(126,285956957.6)(127,283926935.2)(128,281887059.9)
    };
\addlegendentry{\name (No Ticks)}

\addplot[
    color=blue,
    mark=triangle*,
    mark size=2.0pt,
    mark repeat=3,
    ]
    coordinates {
(1,569441975.5)(2,568804310.5)(3,569306862.7)(4,569410518.6)(5,569819397.6)(6,569534654.8)(7,569582129.2)(8,568689663.6)(9,568283081)(10,568762072.6)(11,569108993.2)(12,568638914.1)(13,568726726)(14,569038063.4)(15,568785807.6)(16,567867145.7)(17,568150704.6)(18,568088894.1)(19,568079240.9)(20,567815863.9)(21,567913816.6)(22,567994985.7)(23,568180436.4)(24,567731090.3)(25,567470849.6)(26,567320537.7)(27,567646289.5)(28,567350886)(29,567423116.7)(30,567719744.3)(31,568143443.4)(32,567418408.5)(33,567172871.3)(34,567306051.6)(35,567224537.2)(36,567261301.8)(37,567536879.8)(38,567452208)(39,566993828.1)(40,568156855.9)(41,567344096.6)(42,566464628.6)(43,566439063)(44,565671295)(45,563334697.8)(46,559941314.6)(47,556452134.1)(48,553166035.1)(49,549994796.5)(50,546511145.7)(51,542946195)(52,540062395)(53,537678277.9)(54,534027583.6)(55,531215524.1)(56,527986939.1)(57,525156713.7)(58,522309868.8)(59,519719812.6)(60,517112664.9)(61,514638414.4)(62,512503516.8)(63,509167263.1)(64,506278778.4)(65,498940827.3)(66,492908613)(67,486285155)(68,479951025.8)(69,473727303.2)(70,467205020)(71,461549394.5)(72,455955332.6)(73,450300800.2)(74,445021955.1)(75,439760706.6)(76,434465612.6)(77,429437551)(78,424577828)(79,419859150.1)(80,415152352.8)(81,410824475)(82,406487207.4)(83,402048957.4)(84,397883890.9)(85,393467384.1)(86,389675904.8)(87,385692391.4)(88,381895340.6)(89,378225843.5)(90,374842585)(91,371215344.7)(92,367587673)(93,364100366.6)(94,360695206.2)(95,357567095.6)(96,354424927.1)(97,351257291.7)(98,347904068)(99,344995451.4)(100,342043174)(101,339272338.1)(102,336445866.8)(103,333782947.3)(104,331042913.2)(105,328395175.1)(106,325630376.6)(107,322936779.5)(108,320485657.7)(109,318136899.8)(110,315830670.7)(111,313362975.4)(112,310821077.1)(113,308429417.2)(114,306172794.6)(115,303864108.9)(116,301543474.4)(117,299523202.1)(118,297489841.4)(119,295353904.9)(120,293219502.3)(121,291121546.2)(122,288969420.6)(123,286997616.4)(124,284906407)(125,282940483.3)(126,281118765)(127,279008471.3)(128,277054842.4)
    };
\addlegendentry{On-Host (Ticks)}

\end{axis}
\end{tikzpicture}}  
\caption{For each scenario, there are two VMs with 128 vCPUs each, scheduled on a 128 core socket.}
\label{fig:offload-gce-tickless}
\vspace{0.2cm}
\end{subfigure}

\begin{subfigure}[t]{0.6\linewidth}
\resizebox{\linewidth}{!}{\begin{tikzpicture}
\begin{axis}[
    title={},
    xlabel={\# vCPUs Running \texttt{busy\_loop}},
    ylabel style={align=center}, ylabel={\% Improvement\\of \name},
    xmin=0, xmax=128,
    ymin=0, ymax=12,
    xtick={0, 16, 32, 48, 64, 80, 96, 112, 128},
    ytick={2, 4, 6, 8, 10, 12},
    hide scale,
    ymajorgrids=true,
    grid style=dashed,
    width = \figwidth,
    height = \figimprovementheight,
    legend style={at={(0.309,0.385)},anchor=north},
    every axis plot/.append style={very thick}
]

\addplot[
    color=purple,
    mark=none,
    ]
    coordinates {
(1,11.19692221)(2,11.5132445)(3,11.47063785)(4,11.21362513)(5,11.16616987)(6,11.2873569)(7,11.14980318)(8,11.30080075)(9,11.7247288)(10,11.27867038)(11,11.24796525)(12,11.41842377)(13,11.14524124)(14,11.30977721)(15,11.47887995)(16,11.42496682)(17,11.26763463)(18,11.64954686)(19,11.06361683)(20,11.30097675)(21,11.61717533)(22,11.39172165)(23,11.13410907)(24,11.62559872)(25,11.31365838)(26,10.98096798)(27,11.54628885)(28,11.3772386)(29,10.84333354)(30,10.97192097)(31,9.739908846)(32,6.239595874)(33,4.876426954)(34,4.884617632)(35,5.583010059)(36,4.924898111)(37,4.753709495)(38,5.485664113)(39,5.177155852)(40,4.372381633)(41,5.181295326)(42,5.458071324)(43,4.763172576)(44,5.423566935)(45,5.945299061)(46,5.910756893)(47,6.458842631)(48,6.091591999)(49,5.40708362)(50,5.712979799)(51,5.983762556)(52,5.506070541)(53,5.497448178)(54,5.708129174)(55,5.327746577)(56,5.363248454)(57,5.498635789)(58,5.350123439)(59,5.218944722)(60,5.177978665)(61,4.987070442)(62,4.763948792)(63,4.785227685)(64,4.794426149)(65,4.756994621)(66,4.494003799)(67,4.49985625)(68,4.493294587)(69,4.436018595)(70,4.512541173)(71,4.378727809)(72,4.294403975)(73,4.246850607)(74,4.21767335)(75,4.171144499)(76,4.179637034)(77,4.159025083)(78,4.039213111)(79,3.98984753)(80,3.953859263)(81,3.871857392)(82,3.83523107)(83,3.816450377)(84,3.769595147)(85,3.774726999)(86,3.613369103)(87,3.610980627)(88,3.558711676)(89,3.540042235)(90,3.444212067)(91,3.378600535)(92,3.334565301)(93,3.29625737)(94,3.238445311)(95,3.158446965)(96,3.121689835)(97,3.090423862)(98,3.122431658)(99,3.032356449)(100,2.97810398)(101,2.871772616)(102,2.799768236)(103,2.756778118)(104,2.766154785)(105,2.746417259)(106,2.677964686)(107,2.670535868)(108,2.516270704)(109,2.437919341)(110,2.344895329)(111,2.371453403)(112,2.380553839)(113,2.36431874)(114,2.296382186)(115,2.258917106)(116,2.212990931)(117,2.136944135)(118,2.048098396)(119,2.002938986)(120,1.987158004)(121,1.930671334)(122,1.945394366)(123,1.855151487)(124,1.774821708)(125,1.733955357)(126,1.721049319)(127,1.762836765)(128,1.744137558)
    };

\end{axis}
\end{tikzpicture}}
\caption{The \% improvement of \name (No Ticks) over On-Host ghOSt (Ticks).}
\label{fig:offload-gce-improvement}
\end{subfigure}
\vspace{-1.0em}
\caption{Virtual machine compute performance when scheduled by \name (no timer ticks) vs. on-host ghOSt (ticks).}
\Description[Virtual machine compute performance in Wave.]{Virtual machine compute performance when scheduled by Wave (no timer ticks) vs. on-host ghOSt (ticks).}
\label{fig:gce}
\end{figure}

\textit{Results.} In Figure~\ref{fig:gce}, we run \texttt{busy\_loop} on one vCPU in each VM with all other vCPUs idle.
We spawn additional instances of \texttt{busy\_loop}, one per vCPU, and run the utility on the first hyperthread of all physical cores before using neighboring hyperthreads.
\Cref{fig:offload-gce-tickless} compares the work output of both the \name scheduler and the on-host baseline, and \Cref{fig:offload-gce-improvement} shows the percentage improvement achieved by the \name scheduler.
The \name scheduler shows an 11.2\% improvement over on-host when 1 vCPU is active, 9.7\% when 31 vCPUs are active, and 1.7\% when 128 vCPUs are active.
When many vCPUs are idle, active vCPUs save the tick overhead and experience a turbo boost.
Once 32 vCPUs are active, AMD's turbo governor applies a smaller turbo boost.
Once 64 vCPUs are active, the first sibling on each physical core is busy, and we start running \texttt{busy\_loop} on the second siblings.
When 128 vCPUs are active, no vCPUs receive a turbo boost, so the 1.7\% improvement is solely timer tick overhead savings.
Using one SmartNIC core for scheduling saves $1.7\% * 256 = 4.4$ cores per host; at fleet scale, even a small improvement like this creates significant cost savings.

\subsection{RPC Stack}
\label{sec:evaluation:networking}

The SmartNIC is the entry point for all incoming packets, so it can make better packet steering and thread scheduling decisions that \textit{save host compute and improve cache locality and tail latency}.
We offload both the packet-to-host-core steering policy for Stubby~\cite{stubby} (\S\ref{sec:wave-subsystems:network-rpcs}) and the thread scheduler (\S\ref{sec:wave-subsystems:kernel-thread-scheduling}) to the SmartNIC, co-locating them for improved coordination.
The RPC data plane has already been offloaded in prior work~\cite{dagger, flextoe, 1rma, erss, azure-accelerated-networking, lynx, nanopu}, so we retain this design while additionally offloading the packet-to-host-core steering policy.

\subsubsection{RocksDB Evaluation}

RocksDB runs on the host and a load generator under the same top-of-rack switch serves the system with RocksDB RPC requests, with 99.5\% of requests as 10$\mu$s GETs and 0.5\% as 10ms RANGE queries.
The RPC stack and RocksDB run in different processes and pass requests/replies via shared memory, i.e., host DRAM when the RPC stack is on-host and PCIe MMIO when offloaded. 

\textbf{Comparison Scenarios.} We compare three scenarios:
\begin{enumerate}
\item \textbf{\textit{OnHost-All.}} The ghOSt scheduler and the RPC stack are both on host.
All communication is via host shared memory.
The RPC stack uses 8 cores, the scheduler one, and RocksDB 15.
ghOSt performance degrades when it is located in a different CCX than RocksDB, so we co-locate ghOSt and RocksDB in the same two CCXs while the RPC stack runs in a third CCX.

\item \textbf{\textit{OnHost-Scheduler.}} The scheduler runs on the host while the RPC stack is offloaded to the SmartNIC with \name.
The scheduler and RPC stack communicate via MMIO so that the scheduler knows about incoming RPCs, and RocksDB and the RPC stack also use MMIO communication.
The 8 host cores used above for RPCs are now free.
ghOSt uses 1 core, and RocksDB uses 15.

\item \textbf{\textit{Offload-All.}} The scheduler and RPC stack both run on the SmartNIC with \name, communicating via SmartNIC DRAM.
RocksDB uses all 16 host cores and MMIO for cross-PCIe communication with the RPC stack.
\end{enumerate}

\textbf{Scheduler-RPC Synergy.}
We run the \textit{single-queue Shinjuku policy}~\cite{shinjuku} and preempt threads that exceed a 30$\mu$s time slice.
This ensures that in a highly dispersive workload, short requests do not get stuck behind long ones.
\begin{figure}
\newcommand\fighalfpagewidth{0.48\textwidth}
\newcommand\figthirdpagewidth{0.29\textwidth}
\newcommand\figspacing{2mm}
\newcommand\figwidth{6.5cm}
\newcommand\figstubbyheight{4cm}

\begin{subfigure}[t]{0.6\linewidth}
\resizebox{\linewidth}{!}{\begin{tikzpicture}
\begin{axis}[
    title={},
    xlabel={RocksDB ~Throughput (x1K req/s)},
    ylabel style={align=center}, ylabel={99\% Latency ($\mu$s)},
    xmin=0, xmax=240000,
    ymin=0, ymax=1000,
    scaled x ticks=base 10:-3,
    hide scale,
    ymajorgrids=true,
    grid style=dashed,
    width = \figwidth,
    height = \figstubbyheight,
    legend style={at={(0.595,1.61)},anchor=north},
    every axis plot/.append style={very thick}
]

\addplot[
    color=red,
    mark=square,
    ]
    coordinates {
(24955.0931,185.5)(49926.08359,199)(74866.1773,213)(99852.64327,227)(124801.2648,236)(149723.2857,245)(159629.8704,285)(164649.3914,300.5)(169607.8972,444)(174620.0209,691.5)(179616.4379,1043.5)(184641.5655,1252.5)
    };
\addlegendentry{(1) OnHost-All}

\addplot[
    color=green,
    mark=x,
    ]
    coordinates {
(10972.32272,238)(14921.03663,262.5)(19978.67099,296)(24909.99111,332.5)(29928.73801,374)(34893.16627,424.5)(39925.45132,491)(44874.64446,611)(49883.59101,929.5)(52355.44807,1920.5)
    };
\addlegendentry{(2) OnHost-Schedule}

\addplot[
    color=blue,
    mark=o,
    ]
    coordinates {
(24920.59075,161.5)(49884.12044,175.5)(74850.099,206)(99796.82083,237)(124615.4959,277)(149682.4565,319.5)(159658.7419,363)(164639.2522,438)(169512.0007,537)(174554.9347,826)(179625.2599,1224.5)(184624.0972,1477)
    };
\addlegendentry{(3) Offload-All}

\end{axis}
\end{tikzpicture}}  
\caption{Single-queue Shinjuku.}
\label{fig:evaluation:rpc-sq}
\vspace{0.2cm}
\end{subfigure}

\begin{subfigure}[t]{0.6\linewidth}
\resizebox{\linewidth}{!}{\begin{tikzpicture}
\begin{axis}[
    title={},
    xlabel={RocksDB ~Throughput (x1K req/s)},
    ylabel style={align=center}, ylabel={99\% Latency ($\mu$s)},
    xmin=0, xmax=240000,
    ymin=0, ymax=1000,
    scaled x ticks=base 10:-3,
    hide scale,
    ymajorgrids=true,
    grid style=dashed,
    width = \figwidth,
    height = \figstubbyheight,
    legend style={at={(0.57,1.0)},anchor=north},
    every axis plot/.append style={very thick}
]

\addplot[
    color=red,
    mark=square,
    ]
    coordinates {
(24917.41631,213.5)(49896.84087,235)(74789.61646,234.5)(99812.70062,249)(124652.1422,250)(149620.8048,248)(159562.2894,246.5)(169678.2311,248)(179606.0733,251)(189613.4054,250)(194639.9043,259)(199661.7803,257)(204473.151,256)(209620.7533,262.5)(214463.2103,263)(219627.6019,279.5)(224507.7457,299)(229463.092,338)(232107.53,1222)
    };
\addlegendentry{(1) OnHost-All}

\addplot[
    color=green,
    mark=x,
    ]
    coordinates {
(10956.26108,246.5)(14997.32088,264)(19946.69983,298.5)(24949.64084,339)(29896.75893,380.5)(34967.60811,430)(39849.12977,489.5)(44904.03512,583.5)(49946.03865,741.5)(52762.97825,1024.5)
    };
\addlegendentry{(2) OnHost-Schedule}

\addplot[
    color=blue,
    mark=o,
    ]
    coordinates {
(24903.12633,187.5)(49959.25665,205.75)(74725.51986,231.5)(99695.96925,259)(124663.6756,344.75)(149684.088,340.5)(159594.9024,341.25)(169649.074,371.25)(179579.497,390)(189628.6286,438.625)(194530.0515,422)(199523.2935,464.75)(204591.0809,517.5)(209538.7116,511.25)(214421.2559,602.25)(219643.9435,591.75)(224205.702,604.25)(226747.0038,993.25)
    };
\addlegendentry{(3) Offload-All}

% Hide the legend
\legend{}

\end{axis}
\end{tikzpicture}}
\caption{Multi-queue Shinjuku using RPC request SLO.}
\label{fig:evaluation:rpc-mq}
\end{subfigure}
\vspace{-1.0em}
\caption{RocksDB performance when running the RPC stack and the scheduler on the host or on the SmartNIC.}
\Description[RocksDB RPC performance in Wave.]{RocksDB performance in Wave when running the RPC stack and the scheduler on the host or on the SmartNIC.}
\label{fig:evaluation:rpc}
\vspace{-1.0em}
\end{figure}
\Cref{fig:evaluation:rpc-sq} shows that by running the RPC stack and scheduling on the same node (host or SmartNIC), OnHost-All and Offload-All achieve about identical performance, though Offload-All recovers 9 host cores.
Although Offload-All dedicates one more host core to RocksDB, RocksDB suffers from PCIe overhead, and so performs comparably to OnHost-All.
The offloaded RPC stack does not currently leverage prestaging and prefetching, and we attribute the latency gap to this.
When the RPC stack alone is offloaded in OnHost-Schedule, the on-host scheduler must access the RPC headers via MMIO loads to schedule threads, saturating at a much lower throughput.

\textbf{Results of Apples-to-Apples Comparison.}
RocksDB has one more host core in Offload-All (16 cores) than OnHost-All (15).
In this experiment, throughput is linear with the number of host cores as the remaining resources on host and the SmartNIC are not saturated. The throughput with 15 cores is X*(15/16), so
when Offload-All restricts RocksDB to 15 on-host cores, it performs 6.3\% worse than OnHost-All.

\subsubsection {Leveraging Network Insights}
We port the \textit{multi-queue Shinjuku policy}, which leverages RPC-specific information to provide better performance isolation between SLO classes.
Each RPC request includes an SLO in its payload, which the RPC stack passes to the scheduler.
The scheduler assigns the request to an idle RocksDB thread and adds the thread to a per-SLO run queue.
The \name agent then assigns threads to idle host cores.
The preemption time slice is 30 $\mu$s.

\textbf{Results of Multi-Queue Shinjuku Policy.}
In Figure~\ref{fig:evaluation:rpc-mq}, Offload-All saturates 20.8\% higher than single-queue Shinjuku because it leverages RPC information on the SmartNIC.
Offload-All saturates within 2.2\% of OnHost-All although the latter uses 9 more host cores.
The 2.2\% gap is due to PCIe overhead.
In OnHost-Schedule, the overhead of reading the SLO (not just the RPC header) via PCIe dominates and leveraging the included SLO has no impact, so the gap widens between OnHost-Schedule and the other scenarios relative to single-queue Shinjuku.
This shows that to offload/accelerate an RPC stack and effectively leverage the included SLOs, it is \textit{essential} to also offload scheduling.

\textbf{Results of Apples-to-Apples Comparison.}
When Offload-All restricts RocksDB to 15 host cores as above, it performs 7.4\% worse than OnHost-All due to PCIe overhead. 

\subsubsection {Faster Interconnects Benefit \name}
PCIe is the bottleneck in \name for $\mu$s-scale software.
Coherent interconnects like CXL \cite{cxl}, UPI \cite{upi}, and NVLink \cite{nvlink} improve performance and eliminate the need for software coherence.

\textbf{Evaluation Background.}
To show that \name systems take advantage of coherent interconnects, we emulate a UPI-attached SmartNIC using the host CPU in one socket, and we use the other socket's CPU as the host.
A UPI link connects both sockets, and we re-implement the optimizations in \S\ref{sec:wave-implementation}.
We use AMD’s HSMP frequency scaling driver to set the turbo boost limit to emulate the slower SmartNIC cores.
The host socket runs at 3.5GHz, and we test three different frequency limits for the emulated SmartNIC socket: 3GHz (our SmartNIC SoC frequency), 2.5GHz and 2GHz.

We compare two scenarios. 
\textit{Offload:} The \name scheduler and RPC stack run on the emulated SmartNIC in one socket and RocksDB runs in the other socket.
\textit{On-Host:} The \name scheduler and RPC stack along with RocksDB all run in the same socket, and we run the system at the default 3.5GHz host frequency.
In both scenarios, RocksDB uses the same number of cores, so this is an apples-to-apples comparison.

\textbf{Results of Apples-to-Apples Comparison.} The slowdowns at saturation in offload relative to on-host are 1.3\% (offload @ 3GHz), 2.5\% (2.5GHz) and 3.5\% (2GHz). At 3GHz, UPI yields an improvement of 0.9\% over \name running on our real PCIe-connected SmartNIC.
\name uses shared memory and PCIe effectively, performing well without hardware cache coherence but benefiting from it when available.

\subsection{Memory Management}
\label{sec:evaluation:memory-management}
We next offload SOL~\cite{sol}, described in \S\ref{sec:wave-subsystems:memory-management}.

\subsubsection{Evaluation Background} The \name SOL agent uses DMA to transfer RocksDB's page table entries to the SmartNIC and page migration decisions to the host.
The RocksDB database has 10 billion key-value pairs and is $\sim$100 GiB.
We group pages into 256 KiB batches (64 x 4 KiB pages), and scan each batch with periods ranging from 600ms, 1.2s, 2.4s, ..., 9.6s.
Once per 38.4s epoch, fast batches are migrated to DRAM and slow batches are swapped to disk.
There is one load generator thread that creates 10$\mu$s GET requests and 14 RocksDB workers, each affined to a unique core.
We pin the load generator and RocksDB threads to host cores in the same two CCX's (8 physical cores per CCX) and schedule them with the default on-host Linux kernel scheduler.
The agent runs either on the host (in different CCXs than the load generator and RocksDB) or on the SmartNIC.
SOL requires significant compute well beyond one core, so we parallelize the policy and dedicate up to 16 cores.
In each trial, the agent uses the same number of SmartNIC cores and host cores.

\subsubsection{Results of Apples-to-Apples Comparison.} The table below shows an apples-to-apples comparison of the per-iteration agent loop duration.
The duration decreases with more cores.
Portions of the SOL policy are serial, so the duration does not decrease linearly with the core count.

\vspace{-0.8em}
\begin{table}[H]
\centering
\resizebox{0.86\columnwidth}{!}{
\begin{tabularx}{\linewidth}{|X|X|X|}
\hline
\textbf{\# Cores}
&
\textbf{\name (ms)}
&
\textbf{On-Host (ms)}
\\
\hline
1
&
1,018
&
623
\\
\hline
2
&
576
&
431
\\
\hline
4
&
437
&
354
\\
\hline
8
&
384
&
322
\\
\hline
16
&
364
&
309
\\
\hline
\end{tabularx}%
}
\label{tab:memory-loop-overhead}
\end{table}
\vspace{-1.0em}

\textbf{The Overheads.}
Transferring the page table entries with DMA for the entire RocksDB address space takes $\sim$1ms.
The entire address space is transferred on the first agent iteration, but subsequent transfers are faster as SOL learns which page batches to scan less frequently.
Transferring page migration decisions to the host with DMA takes <1ms on all iterations as only a subset of pages is migrated.
The agents spend most of their time on policy computation, so the offloaded software's per-iteration duration is higher than on-host because it uses weaker ARM cores rather than x86 host cores.
Despite this higher duration, the offloaded software approaches the 300ms period in the SOL paper~\cite{sol}, making it practical to deploy SOL without sacrificing 16 host cores.

\textbf{Effect on RocksDB.} SOL reduces RocksDB's memory usage from $\sim$102 GiB DRAM at startup to $\sim$21.3 GiB DRAM (79\% reduction) after 3 epochs.
Each GET request is 10$\mu$s, so SOL's impact on RocksDB performance is minimal, with a median GET latency of 12$\mu$s and a tail (99\%) of 31$\mu$s.
\section{Conclusion}
\label{sec:conclusion}

\name demonstrates that system software is a practical and effective workload for mid-tier SmartNIC ARM cores, enabling cloud providers to recover host resources while sacrificing minimal  performance.
By leveraging a tailored host-SmartNIC communication API and optimizing for both high-throughput and low-latency interactions, \name manages the PCIe bottleneck and achieves efficient offload of system software.
Ultimately, \name highlights the potential for rethinking system software placement in modern datacenters, unlocking new opportunities for efficiency and scalability.
\section*{Acknowledgments}

We thank our anonymous reviewers, Thomas Wenisch, Jeff Mogul, Steve Gribble, and our shepherd, Dan Tsafrir, for their helpful feedback. Jack Humphries was partially supported by the NSF Graduate Research Fellowship. Jack Humphries and Christos Kozyrakis were partially  supported by the Stanford Platform Lab and its affiliates, and
by ACE, one of the seven centers in JUMP 2.0, a Semiconductor Research Corporation (SRC) program sponsored by DARPA. 

\bibliographystyle{unsrt}
\bibliography{bibliography}

\begin{thebibliography}{100}

\bibitem{nitro}
{AWS Nitro System}.
\newblock \url{https://aws.amazon.com/ec2/nitro/}.
\newblock Last accessed: 2020-11-29.

\bibitem{bluefield}
{NVIDIA BlueField Networking Platform}.
\newblock \url{https://www.nvidia.com/en-us/networking/products/data-processing-unit/}.
\newblock Last accessed: 2024-10-06.

\bibitem{fungible}
{Fungible}.
\newblock \url{https://www.fungible.com/product/dpu-platform/}.
\newblock Last accessed: 2022-12-06.

\bibitem{pensando}
{AMD Pensando}.
\newblock \url{https://www.amd.com/en/accelerators/pensando}.
\newblock Last accessed: 2022-12-06.

\bibitem{intel-mount-evans}
{Intel® Infrastructure Processing Unit (Intel® IPU) ASIC E2000}.
\newblock \url{https://www.intel.com/content/www/us/en/products/details/network-io/ipu/e2000-asic.html}.
\newblock Last accessed: 2023-08-04.

\bibitem{dagger}
Nikita Lazarev, Shaojie Xiang, Neil Adit, Zhiru Zhang, and Christina Delimitrou.
\newblock Dagger: Efficient and fast rpcs in cloud microservices with near-memory reconfigurable nics.
\newblock In {\em Proceedings of the 26th ACM International Conference on Architectural Support for Programming Languages and Operating Systems}, ASPLOS '21, page 36–51, New York, NY, USA, 2021. Association for Computing Machinery.

\bibitem{flextoe}
Rajath Shashidhara, Tim Stamler, Antoine Kaufmann, and Simon Peter.
\newblock {FlexTOE}: Flexible {TCP} offload with {Fine-Grained} parallelism.
\newblock In {\em 19th USENIX Symposium on Networked Systems Design and Implementation (NSDI 22)}, pages 87--102, Renton, WA, April 2022. USENIX Association.

\bibitem{1rma}
Arjun Singhvi, Aditya Akella, Dan Gibson, Thomas~F. Wenisch, Monica Wong-Chan, Sean Clark, Milo M.~K. Martin, Moray McLaren, Prashant Chandra, Rob Cauble, Hassan M.~G. Wassel, Behnam Montazeri, Simon~L. Sabato, Joel Scherpelz, and Amin Vahdat.
\newblock {1RMA: Re-Envisioning Remote Memory Access for Multi-Tenant Datacenters}.
\newblock In {\em Proceedings of the Annual Conference of the ACM Special Interest Group on Data Communication on the Applications, Technologies, Architectures, and Protocols for Computer Communication}, SIGCOMM '20, page 708–721, Virtual Event, USA, 2020. Association for Computing Machinery.

\bibitem{erss}
Alexander Rucker, Muhammad Shahbaz, Tushar Swamy, and Kunle Olukotun.
\newblock Elastic rss: Co-scheduling packets and cores using programmable nics.
\newblock In {\em Proceedings of the 3rd Asia-Pacific Workshop on Networking 2019}, APNet '19, page 71–77, New York, NY, USA, 2019. Association for Computing Machinery.

\bibitem{azure-accelerated-networking}
Daniel Firestone, Andrew Putnam, Sambhrama Mundkur, Derek Chiou, Alireza Dabagh, Mike Andrewartha, Hari Angepat, Vivek Bhanu, Adrian Caulfield, Eric Chung, Harish~Kumar Chandrappa, Somesh Chaturmohta, Matt Humphrey, Jack Lavier, Norman Lam, Fengfen Liu, Kalin Ovtcharov, Jitu Padhye, Gautham Popuri, Shachar Raindel, Tejas Sapre, Mark Shaw, Gabriel Silva, Madhan Sivakumar, Nisheeth Srivastava, Anshuman Verma, Qasim Zuhair, Deepak Bansal, Doug Burger, Kushagra Vaid, David~A. Maltz, and Albert Greenberg.
\newblock Azure accelerated networking: Smartnics in the public cloud.
\newblock In {\em Proceedings of the 15th USENIX Conference on Networked Systems Design and Implementation}, NSDI'18, page 51–64, USA, 2018. USENIX Association.

\bibitem{lynx}
Maroun Tork, Lina Maudlej, and Mark Silberstein.
\newblock Lynx: A smartnic-driven accelerator-centric architecture for network servers.
\newblock In {\em Proceedings of the Twenty-Fifth International Conference on Architectural Support for Programming Languages and Operating Systems}, ASPLOS '20, page 117–131, New York, NY, USA, 2020. Association for Computing Machinery.

\bibitem{nanopu}
Stephen Ibanez, Alex Mallery, Serhat Arslan, Theo Jepsen, Muhammad Shahbaz, Changhoon Kim, and Nick McKeown.
\newblock {The nanopu: A nanosecond network stack for datacenters}.
\newblock In {\em 15th $\{$USENIX$\}$ Symposium on Operating Systems Design and Implementation ($\{$OSDI$\}$ 21)}, pages 239--256, 2021.

\bibitem{aws-nitro-hotchips-2019}
Anthony Liguori.
\newblock {The Nitro Project – Next Generation AWS Infrastructure}.
\newblock Hot Chips: A Symposium on High Performance Chips, 2019.

\bibitem{intel-mount-evans-joint}
{Intel and Google Cloud jointly launch data center accelerator chip}.
\newblock \url{https://www.datacenterdynamics.com/en/news/intel-and-google-cloud-jointly-launch-data-center-accelerator-chip}.

\bibitem{characterizing-off-path-smartnic}
Xingda Wei, Rongxin Cheng, Yuhan Yang, Rong Chen, and Haibo Chen.
\newblock Characterizing off-path {SmartNIC} for accelerating distributed systems.
\newblock In {\em 17th USENIX Symposium on Operating Systems Design and Implementation (OSDI 23)}, pages 987--1004, Boston, MA, July 2023. USENIX Association.

\bibitem{the-rise-of-smartnics}
Kartik Srinivasan.
\newblock {The Rise of SmartNICs}.
\newblock \url{https://semiengineering.com/the-rise-of-smartnics/}.

\bibitem{on-disadvantages-of-programmable-nics}
Jack Zhao, Miguel Neves, and Israat Haque.
\newblock {On the (dis)Advantages of Programmable NICs for Network Security Services}.
\newblock In {\em {2023 IFIP Networking Conference (IFIP Networking)}}, pages 1--9, 2023.

\bibitem{mind-the-gap}
Jack~Tigar Humphries, Kostis Kaffes, David Mazi\`{e}res, and Christos Kozyrakis.
\newblock Mind the gap: A case for informed request scheduling at the nic.
\newblock In {\em Proceedings of the 18th ACM Workshop on Hot Topics in Networks}, HotNets ’19, page 60–68, New York, NY, USA, 2019. Association for Computing Machinery.

\bibitem{clara}
Yiming Qiu, Jiarong Xing, Kuo-Feng Hsu, Qiao Kang, Ming Liu, Srinivas Narayana, and Ang Chen.
\newblock Automated smartnic offloading insights for network functions.
\newblock In {\em Proceedings of the ACM SIGOPS 28th Symposium on Operating Systems Principles}, SOSP '21, page 772–787, New York, NY, USA, 2021. Association for Computing Machinery.

\bibitem{lognic}
Zerui Guo, Jiaxin Lin, Yuebin Bai, Daehyeok Kim, Michael Swift, Aditya Akella, and Ming Liu.
\newblock Lognic: A high-level performance model for smartnics.
\newblock In {\em Proceedings of the 56th Annual IEEE/ACM International Symposium on Microarchitecture}, MICRO '23, page 916–929, New York, NY, USA, 2023. Association for Computing Machinery.

\bibitem{floem}
Phitchaya~Mangpo Phothilimthana, Ming Liu, Antoine Kaufmann, Simon Peter, Rastislav Bodik, and Thomas Anderson.
\newblock Floem: A programming system for nic-accelerated network applications.
\newblock In {\em 13th {USENIX} Symposium on Operating Systems Design and Implementation ({OSDI} 18)}, pages 663--679, Carlsbad, CA, October 2018. {USENIX} Association.

\bibitem{intel-mount-evans-venture-beat}
{Intel partners with Google to deploy ‘Mount Evans’ ASIC-based IPU}.
\newblock \url{https://venturebeat.com/business/intel-partners-with-google-to-deploy-mount-evans-asic-based-gpu}.

\bibitem{no-aws-operator-access}
{No AWS operator access}.
\newblock \url{https://docs.aws.amazon.com/whitepapers/latest/security-design-of-aws-nitro-system/no-aws-operator-access.html}.

\bibitem{l4-microkernel-lessons}
Gernot Heiser and Kevin Elphinstone.
\newblock L4 microkernels: The lessons from 20 years of research and deployment.
\newblock {\em ACM Trans. Comput. Syst.}, 34(1), April 2016.

\bibitem{redleaf}
Vikram Narayanan, Tianjiao Huang, David Detweiler, Dan Appel, Zhaofeng Li, Gerd Zellweger, and Anton Burtsev.
\newblock {RedLeaf}: Isolation and communication in a safe operating system.
\newblock In {\em 14th USENIX Symposium on Operating Systems Design and Implementation (OSDI 20)}, pages 21--39. USENIX Association, November 2020.

\bibitem{l3-to-sel4}
Kevin Elphinstone and Gernot Heiser.
\newblock From l3 to sel4 what have we learnt in 20 years of l4 microkernels?
\newblock In {\em Proceedings of the Twenty-Fourth ACM Symposium on Operating Systems Principles}, SOSP '13, page 133–150, New York, NY, USA, 2013. Association for Computing Machinery.

\bibitem{microkernel-goes-general}
Haibo Chen, Xie Miao, Ning Jia, Nan Wang, Yu~Li, Nian Liu, Yutao Liu, Fei Wang, Qiang Huang, Kun Li, Hongyang Yang, Hui Wang, Jie Yin, Yu~Peng, and Fengwei Xu.
\newblock Microkernel goes general: Performance and compatibility in the {HongMeng} production microkernel.
\newblock In {\em 18th USENIX Symposium on Operating Systems Design and Implementation (OSDI 24)}, pages 465--485, Santa Clara, CA, July 2024. USENIX Association.

\bibitem{okl4}
Gernot Heiser and Ben Leslie.
\newblock The okl4 microvisor: convergence point of microkernels and hypervisors.
\newblock In {\em Proceedings of the First ACM Asia-Pacific Workshop on Workshop on Systems}, APSys '10, page 19–24, New York, NY, USA, 2010. Association for Computing Machinery.

\bibitem{qnx}
Dan Hildebrand.
\newblock An architectural overview of qnx.
\newblock In {\em Proceedings of the Workshop on Micro-Kernels and Other Kernel Architectures}, page 113–126, USA, 1992. USENIX Association.

\bibitem{theseus}
Kevin Boos, Namitha Liyanage, Ramla Ijaz, and Lin Zhong.
\newblock Theseus: an experiment in operating system structure and state management.
\newblock In {\em 14th USENIX Symposium on Operating Systems Design and Implementation (OSDI 20)}, pages 1--19. USENIX Association, November 2020.

\bibitem{xpc}
Dong Du, Zhichao Hua, Yubin Xia, Binyu Zang, and Haibo Chen.
\newblock Xpc: architectural support for secure and efficient cross process call.
\newblock In {\em Proceedings of the 46th International Symposium on Computer Architecture}, ISCA '19, page 671–684, New York, NY, USA, 2019. Association for Computing Machinery.

\bibitem{fuschia-zircon}
{Zircon}.
\newblock \url{https://fuchsia.dev/fuchsia-src/concepts/kernel}.

\bibitem{harmonizing-microkernels}
Jinyu Gu, Xinyue Wu, Wentai Li, Nian Liu, Zeyu Mi, Yubin Xia, and Haibo Chen.
\newblock Harmonizing performance and isolation in microkernels with efficient intra-kernel isolation and communication.
\newblock In {\em 2020 USENIX Annual Technical Conference (USENIX ATC 20)}, pages 401--417. USENIX Association, July 2020.

\bibitem{minix3}
Jorrit~N. Herder, Herbert Bos, Ben Gras, Philip Homburg, and Andrew~S. Tanenbaum.
\newblock Minix 3: a highly reliable, self-repairing operating system.
\newblock {\em SIGOPS Oper. Syst. Rev.}, 40(3):80–89, July 2006.

\bibitem{pikeos}
Robert Kaiser and Stephan Wagner.
\newblock {Evolution of the PikeOS Microkernel}.
\newblock 02 2007.

\bibitem{sel4}
Gerwin Klein, Kevin Elphinstone, Gernot Heiser, June Andronick, David Cock, Philip Derrin, Dhammika Elkaduwe, Kai Engelhardt, Rafal Kolanski, Michael Norrish, Thomas Sewell, Harvey Tuch, and Simon Winwood.
\newblock sel4: formal verification of an os kernel.
\newblock In {\em Proceedings of the ACM SIGOPS 22nd Symposium on Operating Systems Principles}, SOSP '09, page 207–220, New York, NY, USA, 2009. Association for Computing Machinery.

\bibitem{hydra}
R.~Levin, E.~Cohen, W.~Corwin, F.~Pollack, and W.~Wulf.
\newblock Policy/mechanism separation in hydra.
\newblock {\em SIGOPS Oper. Syst. Rev.}, 9(5):132–140, November 1975.

\bibitem{improving-ipc-by-kernel-design}
Jochen Liedtke.
\newblock Improving ipc by kernel design.
\newblock {\em SIGOPS Oper. Syst. Rev.}, 27(5):175–188, December 1993.

\bibitem{on-microkernel-construction}
J.~Liedtke.
\newblock On micro-kernel construction.
\newblock In {\em Proceedings of the Fifteenth ACM Symposium on Operating Systems Principles}, SOSP '95, page 237–250, New York, NY, USA, 1995. Association for Computing Machinery.

\bibitem{filesystem-semi-microkernel}
Jing Liu, Anthony Rebello, Yifan Dai, Chenhao Ye, Sudarsun Kannan, Andrea~C. Arpaci-Dusseau, and Remzi~H. Arpaci-Dusseau.
\newblock Scale and performance in a filesystem semi-microkernel.
\newblock In {\em Proceedings of the ACM SIGOPS 28th Symposium on Operating Systems Principles}, SOSP '21, page 819–835, New York, NY, USA, 2021. Association for Computing Machinery.

\bibitem{mach}
David~L. Black, David~B. Golub, Daniel~P. Julin, Richard~F. Rashid, Richard~P. Draves, Randall~W. Dean, Alessandro Forin, Joseph Barrera, Hideyuki Tokuda, Gerald Malan, and David Bohman.
\newblock {Microkernel Operating System Architecture and Mach}.
\newblock 06 1991.

\bibitem{barrelfish}
Andrew Baumann, Paul Barham, Pierre-Evariste Dagand, Tim Harris, Rebecca Isaacs, Simon Peter, Timothy Roscoe, Adrian Sch\"{u}pbach, and Akhilesh Singhania.
\newblock The multikernel: a new os architecture for scalable multicore systems.
\newblock In {\em Proceedings of the ACM SIGOPS 22nd Symposium on Operating Systems Principles}, SOSP '09, page 29–44, New York, NY, USA, 2009. Association for Computing Machinery.

\bibitem{barrelfish-dc}
Gerd Zellweger, Simon Gerber, Kornilios Kourtis, and Timothy Roscoe.
\newblock Decoupling cores, kernels, and operating systems.
\newblock In {\em 11th USENIX Symposium on Operating Systems Design and Implementation (OSDI 14)}, pages 17--31, Broomfield, CO, October 2014. USENIX Association.

\bibitem{barrelfish-intel-single-chip}
Simon Peter, Adrian Sch{\"{u}}pbach, Dominik Menzi, and Timothy Roscoe.
\newblock Early experience with the barrelfish {OS} and the single-chip cloud computer.
\newblock In Diana G{\"{o}}hringer, Michael H{\"{u}}bner, and J{\"{u}}rgen Becker, editors, {\em 3rd Many-core Applications Research Community {(MARC)} Symposium. Proceedings of the 3rd {MARC} Symposium, Ettlingen, Germany, July 5-6, 2011}, pages 35--39. {KIT} Scientific Publishing, Karlsruhe, 2011.

\bibitem{barrelfish-cosh}
Andrew Baumann, Chris Hawblitzel, Kornilios Kourtis, Tim Harris, and Timothy Roscoe.
\newblock Cosh: Clear {OS} data sharing in an incoherent world.
\newblock In {\em 2014 Conference on Timely Results in Operating Systems (TRIOS 14)}, Broomfield, CO, October 2014. USENIX Association.

\bibitem{barrelfish-msr}
{ Barrelfish: Exploring a Multicore OS}.
\newblock \url{https://www.microsoft.com/en-us/research/blog/barrelfish-exploring-multicore-os/}.

\bibitem{menzi-masters-thesis}
Dominik Menzi.
\newblock Support for heterogeneous cores for barrelfish.
\newblock Master's thesis, ETH Zurich, 2011.

\bibitem{barrelfish-technical-note-001}
Barrelfish technical note 001.
\newblock Technical report, ETH Zurich, 2013.

\bibitem{barrelfish-technical-note-005}
Barrelfish technical note 005.
\newblock Technical report, ETH Zurich, 2013.

\bibitem{nros}
Ankit Bhardwaj, Chinmay Kulkarni, Reto Achermann, Irina Calciu, Sanidhya Kashyap, Ryan Stutsman, Amy Tai, and Gerd Zellweger.
\newblock Nros: Effective replication and sharing in an operating system.
\newblock In {\em 15th {USENIX} Symposium on Operating Systems Design and Implementation ({OSDI} 21)}, pages 295--312. {USENIX} Association, July 2021.

\bibitem{exokernel}
D.~R. Engler, M.~F. Kaashoek, and J.~O'Toole.
\newblock Exokernel: An operating system architecture for application-level resource management.
\newblock In {\em Proceedings of the Fifteenth ACM Symposium on Operating Systems Principles}, SOSP '95, page 251–266, New York, NY, USA, 1995. Association for Computing Machinery.

\bibitem{ix}
Adam Belay, George Prekas, Ana Klimovic, Samuel Grossman, Christos Kozyrakis, and Edouard Bugnion.
\newblock {IX}: A protected dataplane operating system for high throughput and low latency.
\newblock In {\em 11th {USENIX} Symposium on Operating Systems Design and Implementation ({OSDI} 14)}, pages 49--65, Broomfield, CO, October 2014. {USENIX} Association.

\bibitem{shinjuku}
Kostis Kaffes, Timothy Chong, Jack~Tigar Humphries, Adam Belay, David Mazi{\`e}res, and Christos Kozyrakis.
\newblock Shinjuku: Preemptive scheduling for $\mu$second-scale tail latency.
\newblock In {\em 16th {USENIX} Symposium on Networked Systems Design and Implementation ({NSDI} 19)}, pages 345--360, Boston, MA, February 2019. {USENIX} Association.

\bibitem{shenango}
Amy Ousterhout, Joshua Fried, Jonathan Behrens, Adam Belay, and Hari Balakrishnan.
\newblock Shenango: Achieving high {CPU} efficiency for latency-sensitive datacenter workloads.
\newblock In {\em 16th {USENIX} Symposium on Networked Systems Design and Implementation ({NSDI} 19)}, pages 361--378, Boston, MA, February 2019. {USENIX} Association.

\bibitem{reflex}
Ana Klimovic, Heiner Litz, and Christos Kozyrakis.
\newblock Reflex: Remote flash $\approx$ local flash.
\newblock In {\em Proceedings of the Twenty-Second International Conference on Architectural Support for Programming Languages and Operating Systems}, ASPLOS ’17, page 345–359, New York, NY, USA, 2017. Association for Computing Machinery.

\bibitem{zygos}
George Prekas, Marios Kogias, and Edouard Bugnion.
\newblock Zygos: Achieving low tail latency for microsecond-scale networked tasks.
\newblock In {\em Proceedings of the 26th Symposium on Operating Systems Principles}, SOSP ’17, page 325–341, New York, NY, USA, 2017. Association for Computing Machinery.

\bibitem{caladan}
Joshua Fried, Zhenyuan Ruan, Amy Ousterhout, and Adam Belay.
\newblock Caladan: Mitigating interference at microsecond timescales.
\newblock In {\em 14th {USENIX} Symposium on Operating Systems Design and Implementation ({OSDI} 20)}, pages 281--297. {USENIX} Association, November 2020.

\bibitem{snap}
Michael Marty, Marc de~Kruijf, Jacob Adriaens, Christopher Alfeld, Sean Bauer, Carlo Contavalli, Mike Dalton, Nandita Dukkipati, William~C. Evans, Steve Gribble, Nicholas Kidd, Roman Kononov, Gautam Kumar, Carl Mauer, Emily Musick, Lena Olson, Mike Ryan, Erik Rubow, Kevin Springborn, Paul Turner, Valas Valancius, Xi~Wang, and Amin Vahdat.
\newblock Snap: a microkernel approach to host networking.
\newblock In {\em In ACM SIGOPS 27th Symposium on Operating Systems Principles}, New York, NY, USA, 2019.

\bibitem{persephone}
Max Demoulin, Josh Fried, Isaac Pedisich, Marios Kogias, Boon~Thau Loo, Linh Thi~Xuan Phan, and Irene Zhang.
\newblock When idling is ideal: Optimizing tail-latency for highly-dispersed datacenter workloads with perséphone.
\newblock In {\em SOSP 2021}, October 2021.

\bibitem{demikernel}
Irene Zhang, Amanda Raybuck, Pratyush Patel, Kirk Olynyk, Jacob Nelson, Omar S.~Navarro Leija, Ashlie Martinez, Jing Liu, Anna~Kornfeld Simpson, Sujay Jayakar, Pedro~Henrique Penna, Max Demoulin, Piali Choudhury, and Anirudh Badam.
\newblock The demikernel datapath os architecture for microsecond-scale datacenter systems.
\newblock In {\em Proceedings of the ACM SIGOPS 28th Symposium on Operating Systems Principles}, SOSP '21, page 195–211, New York, NY, USA, 2021. Association for Computing Machinery.

\bibitem{dune}
Adam Belay, Andrea Bittau, Ali Mashtizadeh, David Terei, David Mazi{\`e}res, and Christos Kozyrakis.
\newblock Dune: Safe user-level access to privileged {CPU} features.
\newblock In {\em 10th {USENIX} Symposium on Operating Systems Design and Implementation ({OSDI} 12)}, pages 335--348, Hollywood, CA, October 2012. {USENIX} Association.

\bibitem{ghost}
Jack~Tigar Humphries, Neel Natu, Ashwin Chaugule, Ofir Weisse, Barret Rhoden, Josh Don, Luigi Rizzo, Oleg Rombakh, Paul~Jack Turner, and Christos Kozyrakis.
\newblock ghost: Fast and flexible user-space delegation of linux scheduling.
\newblock In {\em Proceedings of the ACM SIGOPS 28th Symposium on Operating Systems Principles CD-ROM}, page 588–604, New York, NY, USA, 2021.

\bibitem{syrup}
Kostis Kaffes, Jack~Tigar Humphries, David Mazi\`{e}res, and Christos Kozyrakis.
\newblock Syrup: User-defined scheduling across the stack.
\newblock In {\em Proceedings of the ACM SIGOPS 28th Symposium on Operating Systems Principles}, SOSP '21, page 605–620, Virtual Event, Germany, 2021. Association for Computing Machinery.

\bibitem{ccuserspace}
Akshay Narayan, Frank Cangialosi, Prateesh Goyal, Srinivas Narayana, Mohammad Alizadeh, and Hari Balakrishnan.
\newblock {The Case for Moving Congestion Control Out of the Datapath}.
\newblock In {\em Proceedings of the 16th ACM Workshop on Hot Topics in Networks}, HotNets '17, page 101–107, Palo Alto, CA, USA, 2017. Association for Computing Machinery.

\bibitem{userfaultfd}
{userfaultfd}.
\newblock \url{https://man7.org/linux/man-pages/man2/userfaultfd.2.html}.
\newblock Last accessed: 2023-08-03.

\bibitem{fuse}
{libfuse}.
\newblock \url{https://github.com/libfuse/libfuse}.
\newblock Last accessed: 2020-08-15.

\bibitem{terra-incognita}
Vasily Tarasov, Abhishek Gupta, Kumar Sourav, Sagar Trehan, and Erez Zadok.
\newblock Terra incognita: On the practicality of {User-Space} file systems.
\newblock In {\em 7th USENIX Workshop on Hot Topics in Storage and File Systems (HotStorage 15)}, Santa Clara, CA, July 2015. USENIX Association.

\bibitem{uio}
{The Userspace I/O HOWTO}.
\newblock \url{https://www.kernel.org/doc/html/v4.14/driver-api/uio-howto.html}.
\newblock Last accessed: 2020-11-10.

\bibitem{apple-driverkit}
{DriverKit}.
\newblock \url{https://developer.apple.com/documentation/driverkit}.
\newblock Last accessed: 2023-08-04.

\bibitem{windows-umdf}
{Overview of UMDF}.
\newblock \url{https://learn.microsoft.com/en-us/windows-hardware/drivers/wdf/overview-of-the-umdf}.
\newblock Last accessed: 2023-08-04.

\bibitem{pond.asplos23}
Huaicheng Li, Daniel~S. Berger, Lisa Hsu, Daniel Ernst, Pantea Zardoshti, Stanko Novakovic, Monish Shah, Samir Rajadnya, Scott Lee, Ishwar Agarwal, Mark~D. Hill, Marcus Fontoura, and Ricardo Bianchini.
\newblock {Pond: CXL-Based Memory Pooling Systems for Cloud Platforms}.
\newblock In {\em Proceedings of the 28th ACM International Conference on Architectural Support for Programming Languages and Operating Systems (ASPLOS)}, Vancouver, BC Canada, March 2023.

\bibitem{nsight}
Roni Haecki, Radhika~Niranjan Mysore, Lalith Suresh, Gerd Zellweger, Bo~Gan, Timothy Merrifield, Sujata Banerjee, and Timothy Roscoe.
\newblock {How to diagnose nanosecond network latencies in rich end-host stacks}.
\newblock In {\em 19th USENIX Symposium on Networked Systems Design and Implementation (NSDI 22)}, pages 861--877, Renton, WA, April 2022. USENIX Association.

\bibitem{ipipe}
Ming Liu, Tianyi Cui, Henry Schuh, Arvind Krishnamurthy, Simon Peter, and Karan Gupta.
\newblock Offloading distributed applications onto smartnics using ipipe.
\newblock In {\em Proceedings of the ACM Special Interest Group on Data Communication}, SIGCOMM '19, page 318–333, New York, NY, USA, 2019. Association for Computing Machinery.

\bibitem{rocksdb}
{RocksDB}.
\newblock \url{https://rocksdb.org}.
\newblock Last accessed: 2020-11-27.

\bibitem{userfaultfd-vm}
{mm: userfaultfd: add new UFFDIO\_SIGBUS ioctl}.
\newblock \url{https://lwn.net/ml/linux-kernel/20230511182426.1898675-1-axelrasmussen@google.com}.
\newblock Last accessed: 2023-08-03.

\bibitem{yelam25pageflex}
Anil Yelam, Kan Wu, Zhiyuan Guo, Suli Yang, Rajath Shashidhara, Wei Xu, Stanko Novakovi\'{c}, Alex~C. Snoeren, and Kimberly Keeton.
\newblock {PageFlex: Flexible and Efficient User-space Delegation of Linux Paging Policies with eBPF}.
\newblock In {\em {USENIX} Annual Technical Conference ({ATC} '25)}. {USENIX} Association, 2025.

\bibitem{vfio}
{ VFIO: Virtual Function I/O}.
\newblock \url{https://www.kernel.org/doc/html/v5.9/driver-api/vfio.html}.

\bibitem{arm-big-little}
{big.LITTLE - ARM}.
\newblock \url{https://www.arm.com/why-arm/technologies/big-little}.
\newblock Last accessed: 2020-11-27.

\bibitem{apple-m-series}
{M1 Overview}.
\newblock \url{https://www.apple.com/ua/business/mac/pdf/Apple-at-Work-M1-Overview.pdf}.
\newblock Last accessed: 2023-08-09.

\bibitem{profiling-a-warehouse-scale-computer}
Svilen Kanev, Juan~Pablo Darago, Kim Hazelwood, Parthasarathy Ranganathan, Tipp Moseley, Gu-Yeon Wei, and David Brooks.
\newblock Profiling a warehouse-scale computer.
\newblock In {\em Proceedings of the 42nd Annual International Symposium on Computer Architecture}, ISCA '15, page 158–169, New York, NY, USA, 2015. Association for Computing Machinery.

\bibitem{google-ipu}
{The next wave of Google Cloud infrastructure innovation: New C3 VM and Hyperdisk}.
\newblock \url{https://cloud.google.com/blog/products/compute/introducing-c3-machines-with-googles-custom-intel-ipu}.
\newblock Last accessed: 2024-10-06.

\bibitem{sol}
Yawen Wang, Daniel Crankshaw, Neeraja~J. Yadwadkar, Daniel Berger, Christos Kozyrakis, and Ricardo Bianchini.
\newblock Sol: Safe on-node learning in cloud platforms.
\newblock In {\em Proceedings of the 27th ACM International Conference on Architectural Support for Programming Languages and Operating Systems}, ASPLOS '22, page 622–634, New York, NY, USA, 2022. Association for Computing Machinery.

\bibitem{stubby}
{The Production Environment at Google, from the Viewpoint of an SRE }.
\newblock \url{https://sre.google/sre-book/production-environment/}.

\bibitem{grpc}
{gRPC}.
\newblock \url{https://grpc.io}.
\newblock Last accessed: 2023-08-03.

\bibitem{cfs}
{The Linux Completely Fair Scheduler}.
\newblock \url{https://www.kernel.org/doc/html/latest/scheduler/sched-design-CFS.html}.
\newblock Last accessed: 2023-08-09.

\bibitem{eevdf-1}
Jonathan Corbet.
\newblock {An EEVDF CPU scheduler for Linux}.
\newblock \url{https://lwn.net/Articles/925371/}.
\newblock Last accessed: 2024-07-01.

\bibitem{eevdf-2}
Jonathan Corbet.
\newblock {Completing the EEVDF scheduler}.
\newblock \url{https://lwn.net/Articles/969062/}.
\newblock Last accessed: 2024-07-01.

\bibitem{clock}
F.~J. Corbato.
\newblock A paging experiment with the multics system.
\newblock In {\em In Honor of P. M. Morse}, pages 217--228. MIT Press, 1969.
\newblock Also as MIT Project MAC Report MAC-M-384, May 1968.

\bibitem{aifm}
Zhenyuan Ruan, Malte Schwarzkopf, Marcos~K. Aguilera, and Adam Belay.
\newblock {AIFM}: {High-Performance}, {Application-Integrated} far memory.
\newblock In {\em 14th USENIX Symposium on Operating Systems Design and Implementation (OSDI 20)}, pages 315--332. USENIX Association, November 2020.

\bibitem{autonuma}
Jonathan Corbet.
\newblock {AutoNUMA: the other approach to NUMA scheduling}.
\newblock \url{https://lwn.net/Articles/488709/}.

\bibitem{lake}
Henrique Fingler, Isha Tarte, Hangchen Yu, Ariel Szekely, Bodun Hu, Aditya Akella, and Christopher~J. Rossbach.
\newblock Towards a machine learning-assisted kernel with lake.
\newblock In {\em Proceedings of the 28th ACM International Conference on Architectural Support for Programming Languages and Operating Systems, Volume 2}, ASPLOS 2023, page 846–861, New York, NY, USA, 2023. Association for Computing Machinery.

\bibitem{deeprm}
Hongzi Mao, Mohammad Alizadeh, Ishai Menache, and Srikanth Kandula.
\newblock Resource management with deep reinforcement learning.
\newblock In {\em Proceedings of the 15th ACM Workshop on Hot Topics in Networks}, HotNets '16, page 50–56, New York, NY, USA, 2016. Association for Computing Machinery.

\bibitem{decima}
Hongzi Mao, Malte Schwarzkopf, Shaileshh~Bojja Venkatakrishnan, Zili Meng, and Mohammad Alizadeh.
\newblock Learning scheduling algorithms for data processing clusters.
\newblock In {\em Proceedings of the ACM Special Interest Group on Data Communication}, SIGCOMM '19, page 270–288, New York, NY, USA, 2019. Association for Computing Machinery.

\bibitem{autophase}
Qijing Huang, Ameer Haj-Ali, William Moses, John Xiang, Ion Stoica, Krste Asanovic, and John Wawrzynek.
\newblock {AutoPhase: Compiler Phase-Ordering for HLS with Deep Reinforcement Learning}.
\newblock In {\em {2019 IEEE 27th Annual International Symposium on Field-Programmable Custom Computing Machines (FCCM)}}, pages 308--308, 2019.

\bibitem{pensieve}
Hongzi Mao, Ravi Netravali, and Mohammad Alizadeh.
\newblock Neural adaptive video streaming with pensieve.
\newblock In {\em Proceedings of the Conference of the ACM Special Interest Group on Data Communication}, SIGCOMM '17, page 197–210, New York, NY, USA, 2017. Association for Computing Machinery.

\bibitem{thompson-sampling}
William~R. Thompson.
\newblock {ON THE LIKELIHOOD THAT ONE UNKNOWN PROBABILITY EXCEEDS ANOTHER IN VIEW OF THE EVIDENCE OF TWO SAMPLES}.
\newblock {\em Biometrika}, 25:285--294, 1933.

\bibitem{roce}
Chuanxiong Guo, Haitao Wu, Zhong Deng, Gaurav Soni, Jianxi Ye, Jitu Padhye, and Marina Lipshteyn.
\newblock {RDMA over Commodity Ethernet at Scale}.
\newblock In {\em Proceedings of the 2016 ACM SIGCOMM Conference}, SIGCOMM '16, page 202–215, Florianopolis, Brazil, 2016. Association for Computing Machinery.

\bibitem{networksupport}
Radhika Mittal, Alexander Shpiner, Aurojit Panda, Eitan Zahavi, Arvind Krishnamurthy, Sylvia Ratnasamy, and Scott Shenker.
\newblock Revisiting network support for rdma.
\newblock In {\em Proceedings of the 2018 Conference of the ACM Special Interest Group on Data Communication}, SIGCOMM '18, page 313–326, New York, NY, USA, 2018. Association for Computing Machinery.

\bibitem{rss}
Microsoft Corp.
\newblock {Receive Side Scaling}.
\newblock \url{http://msdn.microsoft.com/library/ windows/hardware/ff556942.aspx}, 2018.

\bibitem{understanding-pcie-performance}
Rolf Neugebauer, Gianni Antichi, Jos\'{e}~Fernando Zazo, Yury Audzevich, Sergio L\'{o}pez-Buedo, and Andrew~W. Moore.
\newblock Understanding pcie performance for end host networking.
\newblock In {\em Proceedings of the 2018 Conference of the ACM Special Interest Group on Data Communication}, SIGCOMM '18, page 327–341, New York, NY, USA, 2018. Association for Computing Machinery.

\bibitem{dpdk}
{Data plane development kit}.
\newblock \url{http://www.dpdk.org/}.
\newblock Last accessed: 2019-06-26.

\bibitem{pio}
Anastasiia Ruzhanskaia, Pengcheng Xu, David Cock, and Timothy Roscoe.
\newblock {Rethinking Programmed I/O for Fast Devices, Cheap Cores, and Coherent Interconnects}, 2024.

\bibitem{e3}
Ming Liu, Simon Peter, Arvind Krishnamurthy, and Phitchaya~Mangpo Phothilimthana.
\newblock E3: {Energy-Efficient} microservices on {SmartNIC-Accelerated} servers.
\newblock In {\em 2019 USENIX Annual Technical Conference (USENIX ATC 19)}, pages 363--378, Renton, WA, July 2019. USENIX Association.

\bibitem{clicknp}
Bojie Li, Kun Tan, Layong~(Larry) Luo, Yanqing Peng, Renqian Luo, Ningyi Xu, Yongqiang Xiong, Peng Cheng, and Enhong Chen.
\newblock Clicknp: Highly flexible and high performance network processing with reconfigurable hardware.
\newblock In {\em Proceedings of the 2016 ACM SIGCOMM Conference}, SIGCOMM '16, page 1–14, New York, NY, USA, 2016. Association for Computing Machinery.

\bibitem{memif}
Felix~Xiaozhu Lin and Xu~Liu.
\newblock memif: Towards programming heterogeneous memory asynchronously.
\newblock {\em SIGARCH Comput. Archit. News}, 44(2):369–383, March 2016.

\bibitem{cxl}
{HOME | Compute Express Link}.
\newblock \url{https://www.computeexpresslink.org}.

\bibitem{upi}
{Intel(R) Xeon(R) Processor Scalable Family Technical Overview}.
\newblock \url{https://www.intel.com/content/www/us/en/developer/articles/technical/xeon-processor-scalable-family-technical-overview.html}.
\newblock Last accessed: 2022-12-07.

\bibitem{nvlink}
{NVLink and NVSwitch}.
\newblock \url{https://www.nvidia.com/en-us/data-center/nvlink}.
\newblock Last accessed: 2022-12-07.

\bibitem{neoverse}
Andrea Pellegrini, Nigel Stephens, Magnus Bruce, Yasuo Ishii, Joseph Pusdesris, Abhishek Raja, Chris Abernathy, Jinson Koppanalil, Tushar Ringe, Ashok Tummala, Jamshed Jalal, Mark Werkheiser, and Anitha Kona.
\newblock {The Arm Neoverse N1 Platform: Building Blocks for the Next-Gen Cloud-to-Edge Infrastructure SoC}.
\newblock {\em IEEE Micro}, 40(2):53--62, 2020.

\bibitem{gce}
{Compute Engine | Google Cloud}.
\newblock \url{https://cloud.google.com/products/compute}.
\newblock Last accessed: 2024-08-12.

\bibitem{tableau}
Manohar Vanga, Arpan Gujarati, and Bj\"{o}rn~B. Brandenburg.
\newblock Tableau: A high-throughput and predictable vm scheduler for high-density workloads.
\newblock In {\em Proceedings of the Thirteenth EuroSys Conference}, EuroSys '18, New York, NY, USA, 2018. Association for Computing Machinery.

\end{thebibliography}

\end{document}